# Application of Deep Reinforcement Learning for Intrusion Detection in Internet of Things: A Systematic Review

Saeid Jamshidi[a], Amin Nikanjam[a], Kawser Wazed Nafi[a], Foutse Khomh[a], Rasoul Rasta[b]

[a]SWAT Laboratory, Polytechnique, Montréal, H3T 1J4, Quebec, Canada
[b]Department of Computer Engineering, Science and Research Branch, Islamic Azad University, Tehran, Iran

ARTICLE INFO

ABSTRACT

The Internet of Things (IoT) has significantly expanded the digital landscape, interconnecting an unprecedented array of devices, from home appliances to industrial equipment. This growth enhances functionality, e.g., automation, remote monitoring, and control, and introduces substantial security challenges, especially in defending these devices against cyber threats. Intrusion Detection Systems (IDS) are crucial for securing IoT; however, traditional IDS often struggle to adapt to IoT networks' dynamic and evolving nature and threat patterns. A potential solution is using Deep Reinforcement Learning (DRL) to enhance IDS adaptability, enabling them to learn from and react to their operational environment dynamically.
This systematic review examines the application of DRL to enhance IDS in IoT settings, covering research from the past ten years. This review underscores the state-of-the-art DRL techniques employed to improve adaptive threat detection and real-time security across IoT domains by analyzing various studies. Our findings demonstrate that DRL significantly enhances IDS capabilities by enabling systems to learn and adapt from their operational environment. This adaptability allows IDS to improve threat detection accuracy and minimize false positives, making them more effective in identifying genuine threats while reducing unnecessary alerts. Additionally, this systematic review identifies critical research gaps and future research directions, emphasizing the necessity for more diverse datasets, enhanced reproducibility, and improved integration with emerging IoT technologies. This review aims to foster the development of dynamic and adaptive IDS solutions essential for protecting IoT networks against sophisticated cyber threats.

## 1. Introduction

The Internet of Things (IoT) represents a complex network of interconnected devices and systems, ranging from simple home appliances (e.g., thermostats and TVs) to more complex infrastructures such as traffic lights and industrial equipment [1]. Such networks generate substantial data, which can be exchanged among diverse devices, enabling them to communicate and interact in a unified ecosystem [2]. This integration aims to create a seamless network where devices have the flexibility to interact freely and efficiently.
As IoT expands, securing interconnected devices becomes crucial to ensure effective operation across various environments. Intrusion Detection Systems (IDS) play a pivotal role in this context. An IDS is designed to detect and mitigate malicious activities or attack traffic on the network, thereby ensuring the integrity and security of data across the IoT framework [3] [4]. Among the diverse branches of Machine Learning (ML), Reinforcement Learning (RL) and its extension, Deep Reinforcement Learning (DRL), are recognized as potent techniques for enhancing system adaptability and performance [5] [6]. These techniques enable systems to learn and to improve their performance based on experience without requiring explicit programming [7]. RL is primarily concerned with sequential decision-making, which models the environment's dynamics to select actions that maximize the reward. DRL extends RL by incorporating deep neural networks to handle large volumes of data, including unstructured one [8] [9].
The heterogeneous and unpredictable nature of IoT systems underscores the significant challenges faced by traditional IDS. These systems must be adaptable to the variability and volatility of the IoT and capable of processing large volumes of data from diverse IoT sources efficiently. In the context of IoT, DRL is especially effective for enhancing IDS to learn from their operational environment and dynamically adapt to new and evolving threats, thereby improving detection accuracy. This capability is crucial for securing IoT networks, which are highly dynamic and prone to a wide range of cyber threats [10] [11]. The use of DRL aims to foster more robust and adaptive IDS mechanisms, thereby ensuring the security and integrity of IoT networks [12] [13]. Figure 1 provides a schematic depiction of a DRL-based IDS designed for IoT applications.
There is no extensive systematic review on ML-based IDS in IoT security, so there is a discernible lack of focused studies on DRL-based IDS. Prior research, including key studies by Chen et al. [14], Rizzardi et al. [15], and Arshad et al. [16], have explored the implementation of DRL in IoT and the design of DRL-based IDS. These studies have identified best practices and challenges but have not delved deeply into the progress of DRL algorithms for cyber threat detection. Therefore, this systematic review spans the period from 2014 to 2024, a critical ten years during which significant technological

jamshidi.saeid@polymtl.ca, amin.nikanjam@polymtl.ca, Kwnafi@yahoo.com, foutse.khomh@polymtl.ca, rasta.rasoul@azad.ac.ir (S.J.A.N.K.W.N.F.K.R. Rasta [b])
ORCID(s):

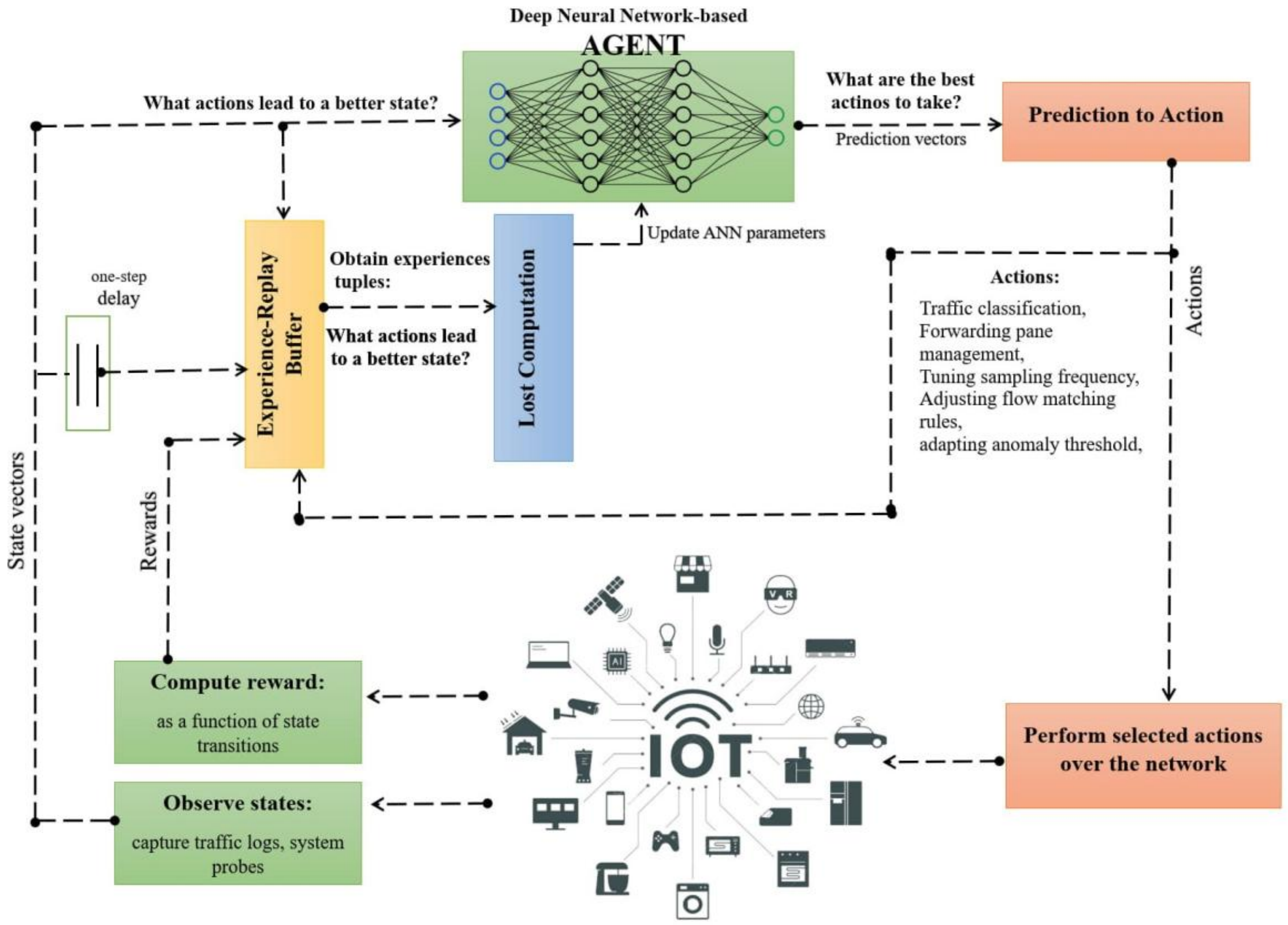

**Figure 1:** Schematic representation of a DRL-based IDS for IoT [15].

breakthroughs have reshaped the landscape of cyber threats, leading to the emergence of sophisticated modern cyber-attacks [17] [18] [19]. This period underscores the potential of DRL in enhancing IDS in IoT systems. It highlights a significant gap in the literature regarding the specialized application of DRL techniques in IoT security, a gap this study aims to address. The systematic review follows the guidelines outlined by Kitchenham and Charters [20], involving stages such as formulating research questions, conducting a literature search, and applying inclusion and exclusion criteria to filter relevant studies. The methodology also includes data extraction and synthesis to identify patterns, trends, and gaps in DRL-based IDS research for IoT. This approach ensures a structured and replicable process for examining the field's current state and identifying future research directions. The significant contributions of this systematic review can be summarized as follows:

- Overview of state-of-the-art DRL in IDS, detailing the evolution and application of these techniques in enhancing IoT security.
- An evaluation of DRL algorithms used in IDS, highlighting their efficiency and application areas.
- An analysis of datasets and benchmarks, discussing their relevance and representation of real-world scenarios for DRL models in IDS.
- Identifying research gaps and challenges provides a roadmap for future investigations in the field.

The remainder of this paper is organized as follows: Section 3 provides an overview of the methodologies employed in this systematic review. In Section 4, we categorize and analyze the state-of-the-art DRL-based IDS approaches. Section 5 examines the datasets used for training and evaluating DRL-based IDS in IoT systems. In Section 6, we present a detailed discussion on the findings of our review, followed by Section 7, which explores the future research opportunities in DRL-based IDS. Finally, Section 8 concludes the paper by summarizing this systematic review's key findings and contributions.

## 2. Background: IoT Security

This section examines IoT architecture, layer-specific security challenges, IDS roles, and mitigation strategies [21]. The IoT connects billions of devices, facilitating data exchange and automation across industries. However, this connectivity introduces significant security challenges, necessitating robust IDS to safeguard IoT ecosystems [22]. IDS mitigates security breaches by analyzing network traffic, device behavior, and system events.

### 2.1. IoT Architecture, Security Challenges, and the Role of IDS

IoT systems consist of three primary layers: the Perception Layer, the Network Layer, and the Application Layer. Each layer faces unique security threats, and IDS plays a vital role in identifying and mitigating these risks [23] [24].

#### 2.1.1. Perception Layer: Physical Interaction and Data Collection

The Perception Layer integrates sensors, actuators, RFID tags, and embedded systems to collect real-world data. As IoT devices operate on limited resources (i.e., lower capacity batteries, lower memory, etc.), they are easily vulnerable to threats, e.g., physical tampering, where attackers manipulate devices to disrupt operations [25]. Mitigation strategies, i.e., tamper-resistant hardware to prevent unauthorized physical access and IDS with real-time anomaly detection to identify irregular network behavior [26], provide essential defenses against cyber threats. Data injection and spoofing, where false data misleads systems, can destabilize infrastructures, e.g., smart grids. Anomaly-based IDS effectively detects such deviations. Firmware exploitation, enabling malware injection, is countered using secure boot and signature-based IDS. Lightweight IDS solutions are crucial for monitoring data streams and maintaining robust security without overburdening devices [27] [28].

#### 2.1.2. Network Layer: Communication Backbone

The Network Layer manages data transmission between devices, gateways, and cloud systems, exposing it to threats targeting communication protocols, traffic flows, and routing mechanisms. Man-in-the-Middle (MITM) attacks intercept and alter data during transmission. Encryption-aware anomaly-based IDS detects unauthorized communication attempts [29] to solve Man-in-the-Middle attacks. Distributed Denial-of-Service (DDoS) attacks, e.g., the Mirai botnet, overwhelm network resources using compromised IoT devices, which can be mitigated through distributed IDS for edge device monitoring and traffic filtering. Routing attacks that manipulate routing tables are countered with signature-based IDS to detect known anomalies and anomaly-based IDS for suspicious patterns. Replay attacks, reusing legitimate packets for unauthorized actions, are mitigated through timestamping and IDS to flag duplicate packets. IDS functionality includes real-time ML traffic monitoring, lightweight IDS at edge nodes, and protocol-specific IDS tuned for MQTT, CoAP, and other IoT protocols to enhance detection accuracy [30] [31].

#### 2.1.3. Application Layer: Data Processing and Service Delivery

The Application Layer processes data for services such as predictive maintenance, remote monitoring, and analytics, interacting with users and external systems. This exposes it to threats, e.g., malware and ransomware, and exploits software vulnerabilities to inject malicious code [32]. Different techniques are already in practice to solve these problems. Among them, Signature-based IDS detects known malware, while behavioral IDS identifies unusual application behavior [33]. Data privacy breaches involving sensitive information, e.g., health or financial data, are mitigated through AES-256 encryption [34] and IDS to detect unauthorized access attempts. Privilege escalation, where attackers exploit weak authentication to gain administrative control, is countered using behavioral IDS to monitor and prevent unauthorized access patterns [35]. IDS at this layer also monitors anomalies, integrates with firewalls and patch management systems, and detects API exploitation through behavior analysis and rule-based monitoring [36].

### 2.2. Cross-Layer Threats and IDS Deployment

Some security threats span multiple IoT layers, requiring advanced IDS. Advanced Persistent Threats (APTs) involve prolonged infiltration using reconnaissance and zero-day vulnerabilities. AI-driven IDS detect these subtle, long-term attack patterns. Insider threats, where authorized users misuse privileges, are addressed with behavioral analytics-based IDS to detect deviations in user activity. Supply chain attacks from compromised hardware or software during manufacturing are mitigated by IDS monitoring hardware integrity and firmware updates for abnormal behavior [37] [38] [39].

### 2.3. Types of IDS in IoT Security

Various IDS technologies are used to address diverse IoT security challenges. Signature-based IDS detect known threats with high accuracy but are ineffective against zero-day exploits. Anomaly-based IDS identifies deviations from normal behavior using ML [40], which is effective against unknown threats but prone to higher false positives. Behavior-based IDS monitors device and network activity to detect unusual patterns, such as unexpected communication [41] [42]. Distributed IDS operate across edge nodes, enabling real-time and localized detection of threats, such as DDoS attacks. This architecture enhances scalability and efficiency. Hybrid IDS combine signature and anomaly-based approaches, leveraging their strengths to detect known and zero-day threats, using signatures for malware and anomaly-based detection for emerging attacks [43].

### 2.4. Advantages of DRL Over Supervised and Unsupervised Learning in IDS

ML is typically categorized into three main types: supervised, unsupervised, and RL. These categories differ in how the ML model processes the data. In supervised learning, the model is provided with data along with corresponding target outputs, and its goal is to learn to predict those outputs based on the input data. Classification tasks involve learning to group data points with the same label into the same category, while regression tasks focus on predicting a continuous value for each input. For example, in autonomous driving, a self-driving car must distinguish between a pedestrian, another car, and a wall. In unsupervised learning, the model does not have labeled data and must identify patterns (i.e., clusters) directly from the raw data. In RL, the model learns by interacting with an environment and receiving feedback as rewards, with no fixed dataset. For example, in autonomous driving, a system responsible for merging into traffic must learn to make safe decisions based on the rewards or penalties it receives from the environment, which reflect the outcomes of its actions. Neural Networks (NN), a type of ML model, have gained popularity in recent years. These networks consist of layers of interconnected neurons, where each neuron is a computational unit that processes information using weights and an activation threshold before passing it onto the next layer. Deep Neural Networks (DNN) are more complex versions of NN, with additional layers and neurons, enabling them to handle more challenging tasks. Deep RL (DRL) is adopting deep neural networks in RL.
DRL offers significant advantages over supervised and unsupervised learning methods in IDS for the IoT. As a key benefit, unlike supervised learning, DRL does not require labeled datasets and can optimize decision-making policies through interaction with the environment. This makes it highly effective for detecting zero-day and unknown attacks [44]. Also, DRL reduces false positive rates compared to unsupervised methods, as it benefits from reward-based learning to distinguish malicious behaviors [45]. Furthermore, DRL dynamically adapts to emerging threats and evolving attack patterns, whereas static training datasets often limit supervised models and struggle with new threats [46]. Another critical advantage of DRL is its ability to optimize resource management in IoT devices, as multi-agent RL techniques help minimize computational overhead and energy consumption. In contrast, supervised models typically require significant processing power [47]. Given these benefits, DRL has been recognized as an efficient approach for enhancing the performance of IDS in IoT networks [48].

## 3. Methodology

This systematic review adhered to the structured methodology for conducting systematic reviews outlined by Kitchenham and Charters [20], which encompasses a series of stages essential for systematic planning, execution, and reporting. The process began with developing pertinent research questions to guide the review. Subsequently, a comprehensive list of keywords was established to facilitate a thorough literature search using a well-defined strategy to capture the most relevant studies. This was followed by defining explicit inclusion and exclusion criteria to ensure a systematic and replicable filtering process. The next step involved outlining a method for data extraction tailored to RQs, ensuring the relevance and systematic organization of the collected data. Finally, the process culminated in the data synthesis, where the extracted information was analyzed and combined to provide a cohesive overview of the research area, identifying key patterns, trends, and gaps in the DRL-based IDS for IoT.

### 3.1. Research questions (RQs)

The primary objective of this systematic review is to examine DRL models in the context of the IoT. This entails a detailed analysis of various DRL models developed specifically for IoT applications, covering the last ten years (Section 4). To systematically address this objective, the following RQs have been formulated:

**RQ1: What is the current state-of-the-art of the DRL-based IDS in IoT?**
This research question examines state-of-the-art DRL-based IDS applications and the most commonly used algorithms in the IoT context.

**RQ2: Which datasets are used for experimental analysis with DRL-based IDS in IoT?**
This research question focuses on the datasets used to train and evaluate DRL models for IDS. It aims to identify the data types that are instrumental in developing these models, with an emphasis on understanding their real-world applicability and effectiveness in detecting intrusions.

**RQ3: What are the current limitations and future research opportunities in DRL-based IDS for IoT?**
This research question aims to identify gaps in the current research and propose areas for future investigation regarding the use and application of DRL in the context of IDS for the IoT infrastructure.

### 3.2. Search strategy

This research utilized automated and manual search strategies to compile a comprehensive list of research papers aligning with the study's objectives. Computerized searches were conducted in several databases e.g., Web of Science to ensure broad coverage of the literature. A manual search was also undertaken, which involved exploring search engines and reviewing the reference lists of related publications to identify relevant studies. The following is an outline of the search process adopted in this systematic review:

- Initially, we selected precise terms based on a thorough review of the relevant literature to ensure both relevance and specificity in the search process. These terms were carefully identified to capture the essential concepts underpinning our study.
- Subsequently, we expanded our search criteria to include related terms and synonyms by utilizing keyword variations. For example, we included terms such as "Intrusion Detection System" along with their acronyms (e.g., "IDS") to encompass a wide array of relevant studies and enhance the comprehensiveness of our literature review.
- The searched terms used in this systematic review had a combination of terms related to the technology and application aspect of the study:
  **("Reinforcement Learning" OR "Deep Reinforcement Learning" OR RL OR DRL) AND ("Internet of Things" OR IoT OR "Intrusion Detection System" OR IDS)**.
  We selected these keywords to ensure comprehensive coverage of the literature on DRL models in the context of IDS, especially in IoT systems.

In conducting this systematic review, the research utilized three significant engineering and scientific databases **Web of Science, Compendex**, and **Inspec**[1] , to collect relevant research papers.

[1] Web of Science: https://www.webofscience.com; Compendex and Inspec: https://www.engineeringvillage.com;

We chose these databases because they cover the most extensive literature in engineering, computer science, and information technology [49] [50]. After utilizing these databases, we searched Google Scholar using relevant keywords. We carefully reviewed the titles and abstracts of papers from the first ten pages of search results, covering publications from 2014 to 2024, to ensure that no existing papers were directly related to our systematic review. This preliminary step helped us confirm the uniqueness of our study and expand our search space by identifying additional relevant papers for further review.

#### *3.2.1. Inclusion and Exclusion Criteria*

In this step, the relevance of the papers was manually inspected by two independent reviewers (the authors). Initially, the reviewers assessed the titles, abstracts, and, where necessary, the full texts of all papers to determine their pertinence to the study. Specific inclusion and exclusion criteria guided this assessment to ensure that only relevant and high-quality research was selected. The following inclusion criteria were applied:

- **Scope of Publication:** Only original research papers published in peer-reviewed journals were considered. The reviewers applied this criterion to ensure the inclusion of high-quality and relevant studies. Peer-reviewed journal publications typically undergo a more rigorous evaluation process than conference papers. While conference papers are also peer-reviewed, they often have shorter review cycles and may present preliminary findings that are further developed in subsequent journal articles. Therefore, conference papers were excluded to focus on comprehensive and fully developed research.

- **Relevance to IDS:** The study must specifically address IDS or their applications in the field, a criterion evaluated by the reviewers during the selection process.

- **Use of DRL Techniques:** The paper must utilize DRL techniques in the context of IDS in the IoT framework. This was another key criterion checked by the reviewers.

- **Publication Period:** The research should have been published between 2014 and 2024, capturing recent advancements in the field.

To maintain the quality and relevance of the systematic review, papers were excluded based on the following criteria:

- **Irrelevant Focus:** Studies that discuss DRL without mentioning its application to IDS and those that mention IDS but do not utilize DRL were excluded. The reviewers ensured this by carefully assessing the focus of each study.

- **Publication Type:** Conference papers, secondary works (e.g., surveys and review papers), research proposals, workshops, letters, and undergraduate theses were excluded. The reviewers applied this exclusion criterion to focus on original research contributions that provide primary data and novel insights rather than summaries or overviews of existing research. Similar SLRs in the field support excluding conference papers, which often prioritize journal articles to ensure a higher standard of research rigor and completeness.

- **Language:** Studies on DRL-based IDS for IoT not written in English were excluded, as determined by the reviewers.

### 3.3. Paper selection

This section describes the procedure for selecting papers and filtering out irrelevant ones.

#### *3.3.1. Duplicated papers*

Following the PRISMA guidelines [51], [52], as depicted in Figure 2, our systematic review process commenced with the initial retrieval of 4,235 papers from various databases, including Compendex, Inspec, and WoS. The process of removing duplicates, facilitated by the Rayyan tool [53], excluded 2,210 duplicates. After removing duplicates, 889 papers were assessed for eligibility based on predefined inclusion and exclusion criteria, such as relevance to DRL-based IDS in IoT, publication date range (2014-2024), and whether the study was peer-reviewed. Studies that did not focus on DRL, IDS, IoT, or those lacking sufficient methodological rigor were excluded. A more detailed appraisal of the remaining 444 papers involved a closer examination of the abstracts and, when necessary, the full texts to ensure they directly addressed the RQs. This resulted in identifying 351 papers relevant to our RQs, especially those that provided empirical data, proposed novel methods, or offered significant insights into DRL-based IDS in IoT. After thorough evaluation and synthesis (Section 3.5), 89 papers were included in the final analysis, from which 36 papers were selected for an in-depth review. This process ensured that only relevant and high-quality research was included in this study.

### 3.4. Analysis of Included Papers

Figure 3 demonstrates the frequency of research contributions in DRL-based IDS in IoT across different countries, based on the selected papers in IoT and the scope of this study. The data shows that researchers from India and China have made the most significant contributions [54], each with a frequency of ten publications. This indicates a higher level of involvement and focus in this field by these two countries, likely driven by their rapid adoption and implementation of IoT technology. Other contributors include Saudi Arabia and the USA, with four publications each, followed by the UK with 3. Several other countries, including Canada, Spain, Hong Kong, Pakistan, Australia, Singapore, and others, show moderate to low contributions, highlighting the global interest and research activity in DRL-based IDS in the IoT domain.

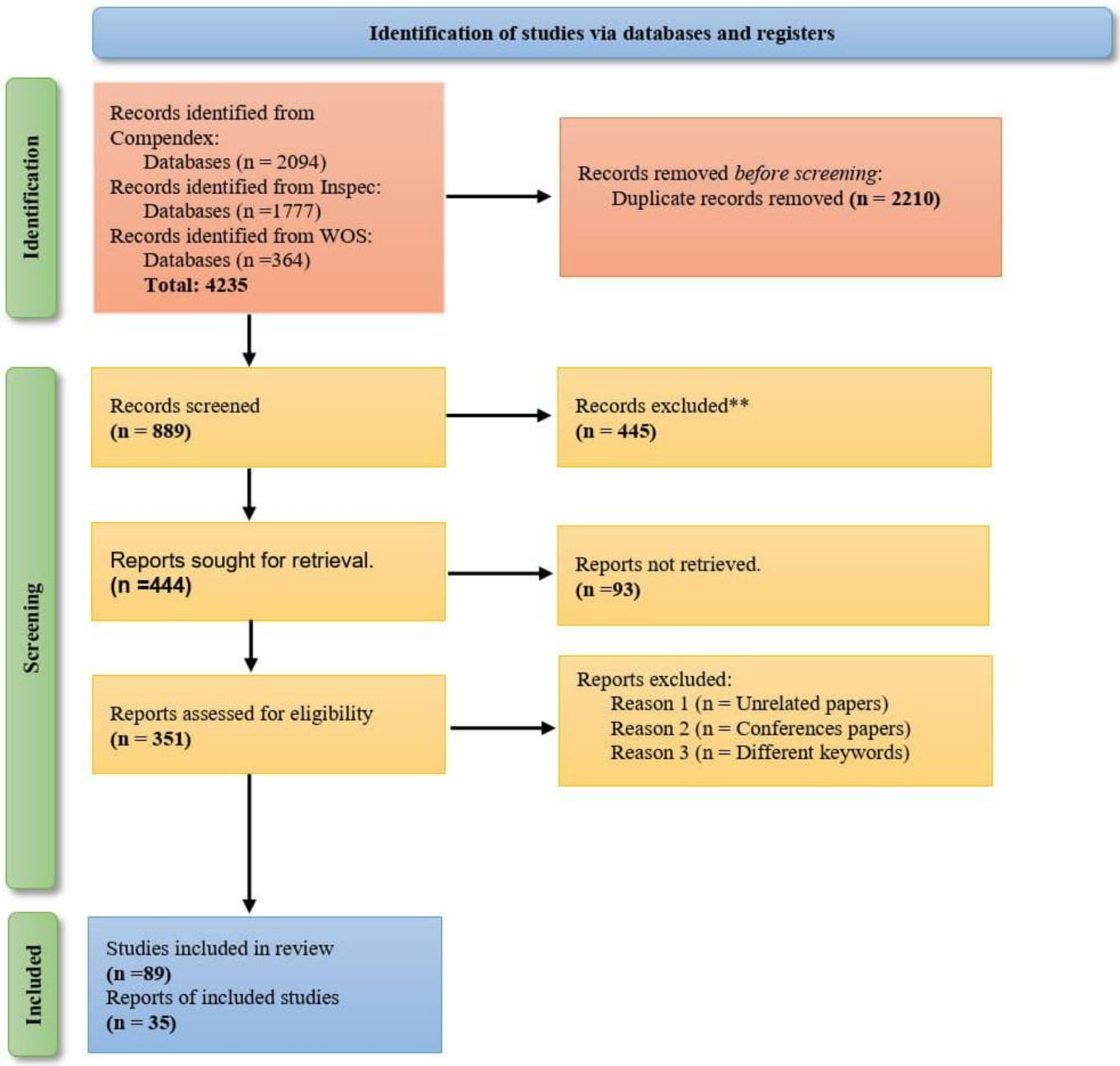

**Figure 2:** PRISMA 2020 Flow Diagram for our systematic reviews.

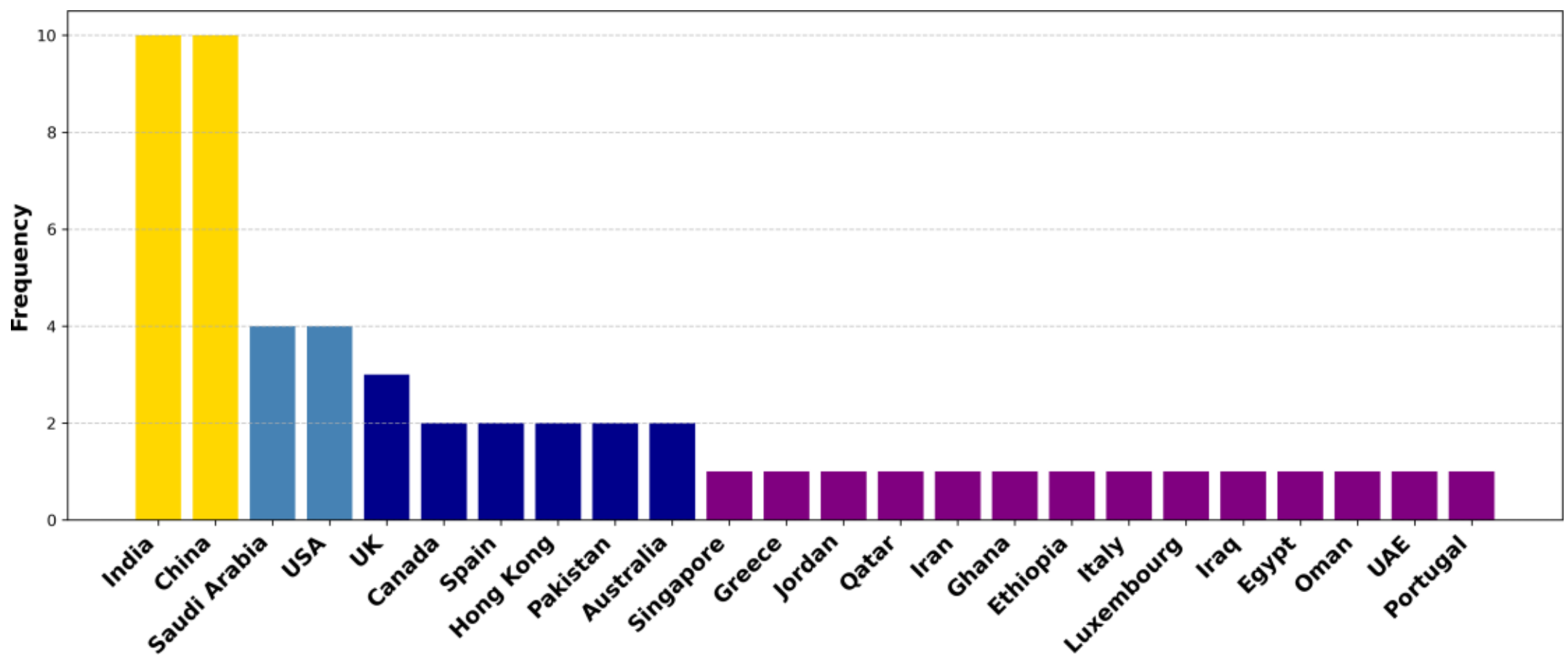

**Figure 3:** Distribution of research papers by country.

Figure 4 presents a keyword network that maps the themes and concepts across the papers included in this systematic review. The graph features 120 keywords extracted from the analyzed papers' titles, abstracts, and keyword lists. The thickness of the lines between pairs of keywords indicates the frequency of their co-occurrence, demonstrating how often these keywords are linked in the literature. The size of the circles and the corresponding font sizes of the keywords indicate their relative prominence in the reviewed studies. Additionally, the color coding represents thematic clustering, with closely related keywords appearing nearer to each other in the same cluster, highlighting the strength of their associations. This mapping also illustrates the interconnections between different thematic clusters. Key terms such as "reinforcement learning," "deep reinforcement learning," "intrusion detection systems," and "security" are notably prominent, reflecting their central role in research related to IoT security from 2014 to 2024.

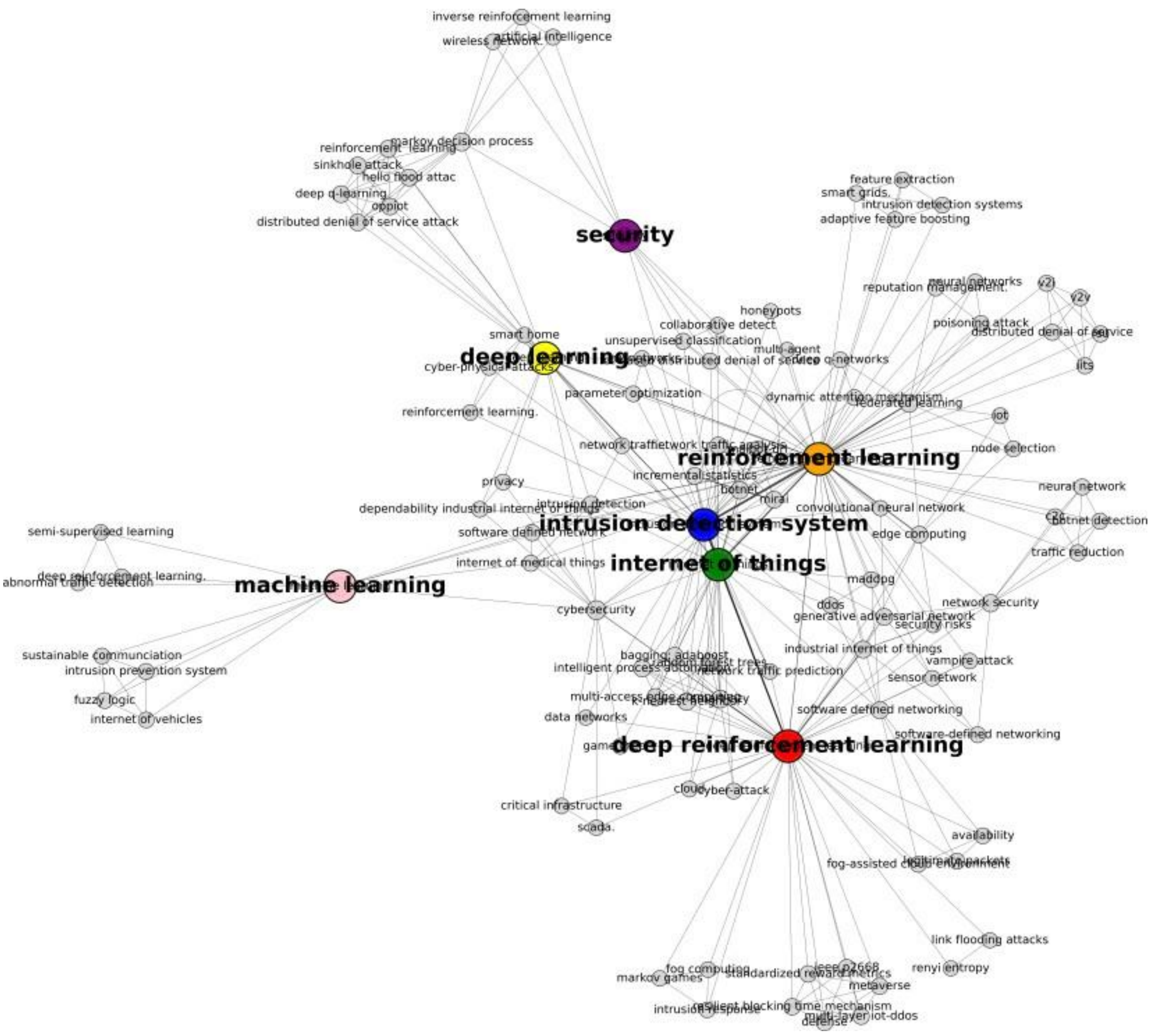

**Figure 4:** The network of the top 120 connected words in papers' titles, abstracts, and keywords.

### 3.5. Data Extraction

We extracted general information and specific details related to the RQs formulated for this systematic review to efficiently analyze the vast amount of information in the IoT set of papers reviewed. This approach ensures that each paper is systematically evaluated for its contributions to DRL-based IDS in IoT.

- General information
- Title
- URL to the paper
- Authors
- Year
- Publication venue

### 3.6. Statistical findings

Figure 5 provides valuable insights into the publishing trends in DRL-based IDS for IoT, helping to identify the key academic and professional venues where significant research is being disseminated. Understanding these trends is crucial for recognizing the authoritative sources that shape the field, allowing researchers to focus on high-impact publications and emerging areas of interest. A significant number of papers, 13 in total, were published under IEEE, highlighting its commitment to advancing research in this domain. Elsevier follows with 11 papers, underscoring its role in disseminating key developments in DRL for IDS. Contributions from Wiley, MDPI, and Springer, with a combined total of 13 papers, further reflect the growing engagement of these publishers in promoting DRL-based security solutions for IoT networks.

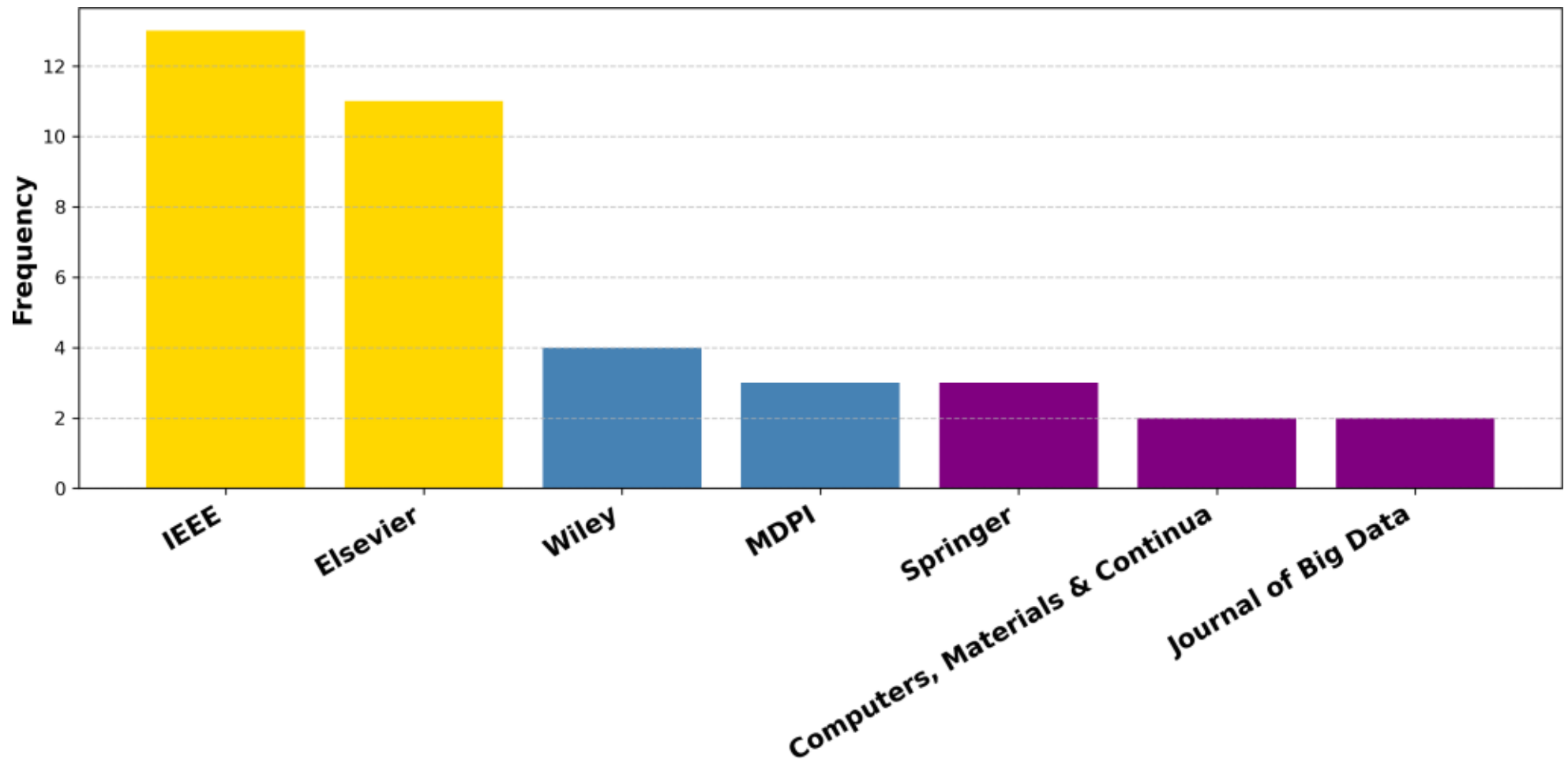

**Figure 5:** Number of papers published per venue.

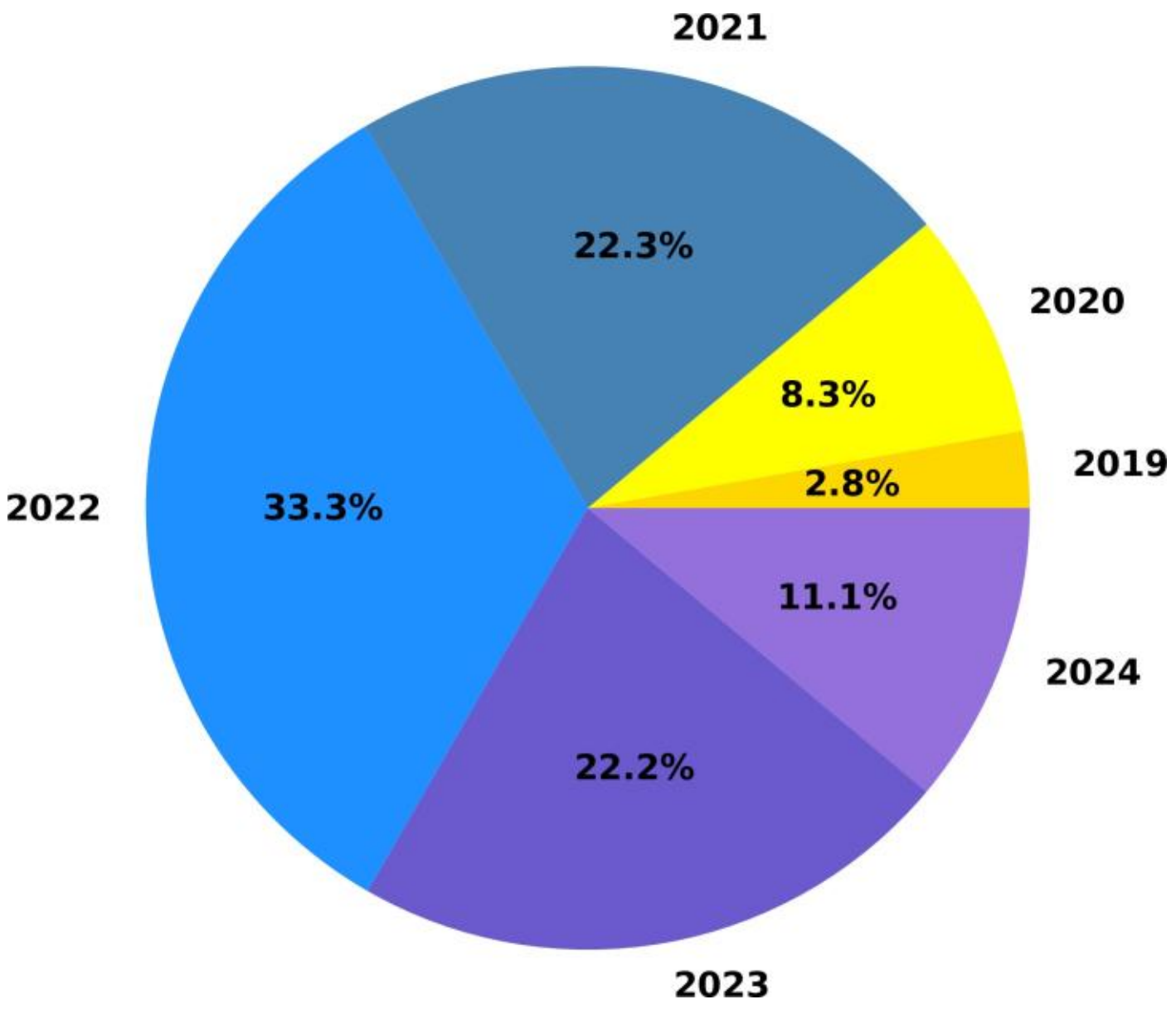

**Figure 6:** Annual trend of research paper publication.

Additionally, contributions from less prominent journals ( e.g., Computers, Materials, and Continua and Journal of Big Data), each with two papers, indicate a developing but growing interest in this research area. This awareness of where cutting-edge advancements are being made and which publishers prioritize this rapidly evolving domain is essential for guiding future research directions and ensuring alignment with the highest standards in the field.
The temporal distribution of research on DRL-based IDS in the IoT, as depicted in Figure 6, reveals a notable increase in scholarly activity. Specifically, 2022 witnessed a significant peak, with one-third (33.3%) of the publications indicating a pivotal year for advancements in this domain. In 2023, the research interest remained strong, accounting for 22.2% of the studies. This was consistent with 2021, which also represented 22.2% of the publications, suggesting a steady focus on the topic. Prior years, such as 2020 and 2019, showed smaller proportions of the research corpus, with 8.3% and 2.8%, respectively, illustrating the emerging stages of DRL-based IDS in IoT. By 2024, the research volume slightly decreased to 11.1%, suggesting a stabilization of interest after the previous surge.

## 4. Categorization of state-of-the-art papers

This section provides a structured categorization of the selected state-of-the-art papers to address RQ1. The types of DRL algorithms organize the categorization. These algorithms represent diverse approaches applied in IDS for IoT systems, as illustrated in Figure 7.

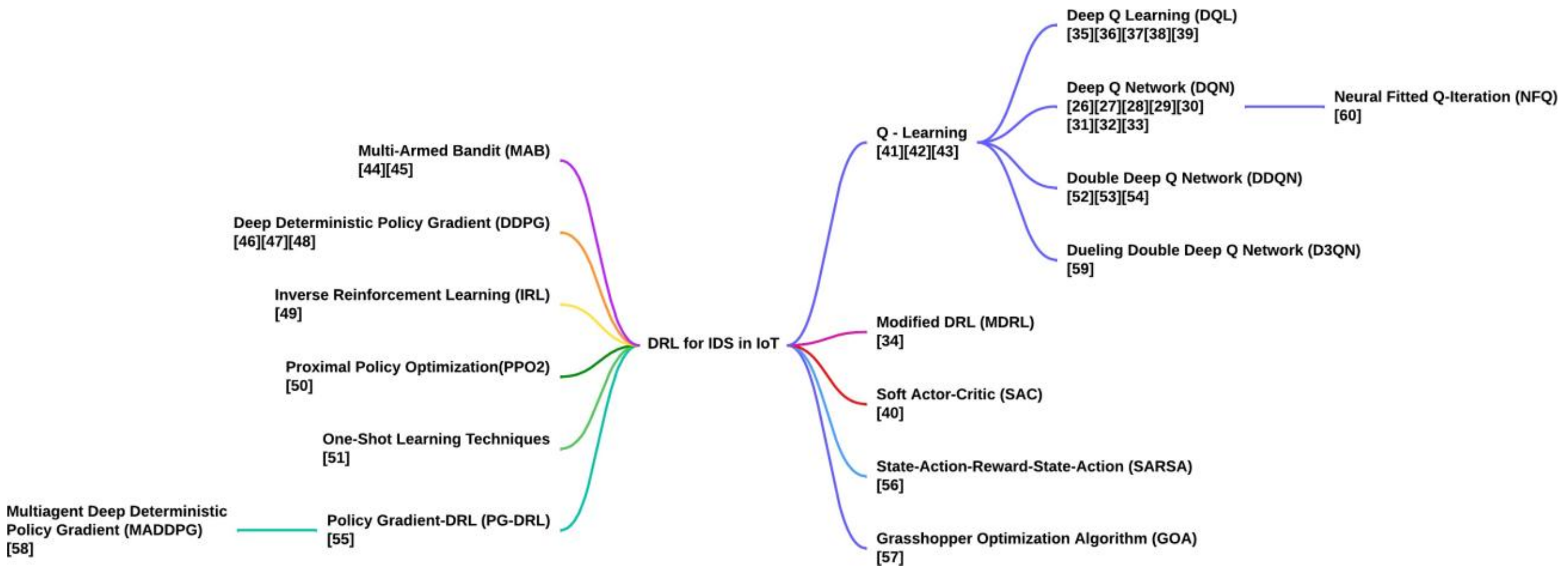

**Figure 7:** Categories and variations of DRL models reported in the literature for IDS in IoT systems.

## 4.1. Deep Q Network (DQN)

Yucheng Liu et al. [55] proposed a multi-layered defense system based on DQN to dynamically adjust the IP blocking time in IoT systems. This system has the advantage of reducing the false positive rate to a great extent and not disrupting the service due to the wrong threat identification. This study achieved an accuracy of 96% in single-layered attack scenarios and 97% in multi-layered attack scenarios. Underlying such practical utility and potential for broader adoption in IoT security is adherence to IEEE P2668 standard, with consequences for scalable applications in the current emerging Metaverse.

Xing Liu et al. [56] explored the application of DRL to the Industrial Internet of Things (IIoT), presenting a system design that evaluates vulnerabilities to adversarial attacks. The study categorized the attacks into two main types: function-based attacks that disrupt the DRL model during training and performance-based attacks that manipulate system operations post-training. Empirical results from simulations demonstrate that these attacks significantly impair system functionality, highlighting the need for robust security measures in IIoT systems.

T.V. Ramana et al. [57] proposed an advanced IDS framework utilizing the RL-DQN model for IoT systems. This dual-layer security system operates across edge and cloud layers, with the edge network achieving a binary attack classification accuracy of 92.8% and the cloud network performing multi-attack classification with an accuracy of 98.2% on the UNSW-NB-15 dataset. The framework employs Markov Decision Processes (MDP) to dynamically adapt to varying network conditions, thereby enhancing the robustness of IDS. The RL-DQN model outperformed traditional ML models, including Random Forest (86.9% accuracy), Gradient Boosting Machine (88.3% accuracy), and Convolutional Neural Networks (90.9% accuracy) on similar datasets. The RL-DQN model demonstrated a recall rate of 98.7% and a precision rate of 93.2%, highlighting its superiority in detecting and classifying intrusions across various datasets, including the IoTID20 dataset.

Sreekanth Vadigi et al. [58] introduced an innovative IDS that leverages Federated Learning (FL) combined with DRL, specifically employing DQN across distributed agents in IoT enterprise networks. The system incorporates a dynamic attention mechanism to enhance detection accuracy while preserving data privacy across the network. Experimental evaluation of the proposed system on the ISOT-CID and NSL-KDD datasets demonstrated impressive results. The system achieved an accuracy of 99.66% on the ISOT-CID dataset and 96.7% on the NSL-KDD dataset, coupled with very low false positive rates of 0.0017 and 0.019, respectively. These performance metrics surpass conventional IDS approaches, such as Support Vector Machines (SVMs) and Decision Trees (DT), which typically exhibit lower accuracy levels, often between 85% and 90%.

Xiaoxue Ma et al. [59] proposed a decision-making framework for intrusion response in fog computing environments, utilizing a Minimax-DQN variant. This adaptation of DQN optimizes strategies to respond to cyber-attacks by modeling the interactions between IDS and attackers as a Markov game. Their model significantly enhances the strategic decision-making capabilities of fog computing environments, effectively demonstrating the practical application of DRL in mitigating cyber risks. The experimental results validate the model's efficacy, showing that the proposed algorithm substantially increases the IDS's probability of winning against attackers, achieving a win rate of 94.6% against random strategies and outperforming standard DQN setups by 3.7% Zhang et al. [60] introduced a defense mechanism for IoT edge devices using DRL in IoT, a zero-sum game framework. They enhance the interaction model between attackers and edge devices by integrating a modified k-nearest Neighbors (kNN) algorithm with Dynamic Time Warping (DTW). This approach analyzes network traffic and data interactions, which is crucial for identifying vulnerabilities. The strategy optimizes defensive responses, adapting to evolving IoT challenges and improving device resilience against cyber threats. Empirical results validate the model's effectiveness, showing decreased attacker payoffs and enhanced security metrics. This leads to a notable reduction in successful attacks, demonstrating the practical utility of their approach in real-world scenarios.

Mohammad Al-Fawa'reh et al. [61] introduced a sophisticated detection system for malware botnets in IoT networks named MalBoT-DRL. The system leverages DRL to manage the evolving patterns of malware dynamically. It incorporates damped incremental statistics, specifically mean, variance, and covariance of network traffic features, along with an attention reward mechanism in IoT. This combination enhances the system's adaptability and generalizability, effectively addressing the issue of model drift. Performance evaluations of MalBoT-DRL, conducted through trace-driven experiments on two representative datasets, demonstrate its efficacy, with the system achieving an average detection accuracy of 99.80% in the early detection phase and 99.40% in the late detection phase.

By adopting traditional DRL frameworks, Manuel Lopez-Martin et al. [62] replace the interactive real-time system with a simulated environment that utilizes pre-recorded datasets of intrusions, specifically NSL-KDD and AWID. This method generates rewards based on detecting errors during training, effectively leveraging supervised learning techniques. The researchers evaluated several DRL models, including DQN (96.87% accuracy), Double Deep Q-Network (DDQN) (97.85% accuracy), Policy Gradient (PG) (94.50% accuracy), and Actor-Critic (AC) (95.12% accuracy), with DDQN emerging as the best-performing model. The study highlighted that DRL can significantly enhance IDS capabilities, offering improvements in both speed and accuracy over conventional ML techniques.

### 4.2. Modified DRL (MDRL)

Almasri et al. [63] suggested a new two-phase security solution to increase the level of safety for IoT devices in smart cities. In the first phase, the system uses a cascaded adaptive neuro-fuzzy inference system to detect anomalies in network traffic from possibly compromised devices, issuing an alarm to system administrators. Subsequently, the MDRL strategy isolates the compromised devices, cutting off their network communication to prevent further damage. This approach achieved a very high accuracy of about 98.7%, with minimal false alarm rates, and hence, guarantees smooth operation and effective delivery of smart city network services.

### 4.3. Deep Q-Learning (DQL)

Jiushuang Wang et al. [64] introduced ReLFA, an advanced IDS specifically designed to counteract Link Flooding Attacks (LFA) in Software-Defined Networking (SDN) for IoT systems. ReLFA utilizes Rényi entropy to detect anomalies in network traffic, effectively identifying the subtle and varied patterns characteristic of LFAs. This IDS leverages DQL to adjust network routing dynamically, ensuring the system maintains optimal performance even under attack conditions. By integrating DRL into the IDS, ReLFA detects and mitigates attacks in real-time, enhancing IoT networks' overall security and resilience. The effectiveness of ReLFA as an IDS is demonstrated through simulation results, showing that ReLFA achieves faster rerouting and mitigation compared to existing methods, such as LFADefender and Woodpecker. Specifically, ReLFA's rerouting process is much quicker, reducing the time required to alleviate the impact of LFAs and restoring normal network operations more efficiently.

Nisha Kandhoul et al. [65] proposed the Deep Q-Learning Security (DQNSec), a novel routing protocol for the Opportunistic Internet of Things (OppIoT) that leverages DQL. This protocol conceptualizes OppIoT as a Markov decision process and employs value-based and policy-based DQL methods to predict and mitigate threats dynamically. Their extensive simulations, using real data traces from Kaggle's Microsoft Malicious dataset, underscore the robustness of DQNSec against traditional ML-based routing protocols. Specifically, the DQNSec protocol showcased a delivery probability of 47.9%, which was significantly superior to other ML-based solutions such as RFCSec (35.4%), RLProph (33.4%), CAML (32%), and MLProph (29.4%). Additionally, the protocol demonstrated lower average latency and packet drop rates, with an average latency of 1874.58 seconds and 2679 packets dropped, establishing its efficiency, accuracy, and enhanced responsiveness to network anomalies.

Jiadai Wang et al. [66] explored the enhancement of security in the SDN-enabled IIoT through DRL. Their study highlighted that Forwarding Nodes (FNs) are vulnerable to a spectrum of cyber-attacks. They proposed a novel attack tolerance scheme that employs DQL to adaptively direct traffic away from compromised FNs, thereby minimizing the impact of these attacks. The scheme's effectiveness is bolstered by using a Generative Adversarial Network (GAN) to generate realistic network traffic data, enhancing both the training and the evaluation of the DRL model. This methodology not only improved the success rate of IIoT traffic reaching its destination and achieving near-optimal performance but also strengthened the overall security framework of IIoT against diverse attack vectors.

Kezhou Ren et al. [67] introduced MAFSIDS, a sophisticated IDS utilizing DQL to enhance the effectiveness of feature selection in network security systems. By incorporating a Graph Convolutional Network (GCN), MAFSIDS selectively processes network features, achieving high accuracies of 96.8% and 99.1% on CSE-CIC-IDS2018 and NSL-KDD datasets, respectively. This approach significantly minimizes feature redundancy, reducing the original feature set by approximately 80%. Integrating DQL with GCN not only refines the feature selection process but also optimizes the overall performance of IDS, streamlining the identification and classification of network threats.

Bin Yang et al. [68] investigated the application of DRL, specifically DQL and Policy Gradient methods, in their advanced network IDS operating at both packet and flow levels. These DRL algorithms enhance the system's ability to detect and manage cyber threats efficiently across these dimensions: packet-level, which involves the analysis of individual data packets, and flow-level, where the traffic flow in IoTthe network is examined. Utilizing the CICDDoS2019 dataset, the research demonstrated significant improvements in detection performance, achieving an accuracy rate of up to 98.78%, with a marked increase in detection speed and adaptability to new, unseen cyber threats.

### 4.4. Soft Actor-Critic (SAC)

Yuming Feng et al. [69] explored deploying the SAC model in IoTDRL to enhance DDoS attack detection in IoT systems. Their approach innovatively adapts the parameters of an unsupervised classifier at IoT edge gateways, which are especially vulnerable to abrupt and extensive variations in network conditions. In this scenario, traditional adaptive methods falter. The effectiveness of the SAC model in this context is underscored by its ability to respond to environmental changes dynamically across various IoT devices. This method expedites detection and optimizes resource utilization by fostering a collaborative network among edge devices sharing data and detection insights. This is crucial for strengthening the overall security infrastructure of IoT networks against sophisticated and large-scale DDoS attacks. Significantly, the implementation of this approach has yielded a detection accuracy exceeding 95% in real-world scenarios, demonstrating its significant potential to detect even stealthy DDoS attacks effectively.

### 4.5. Q-learning

H. Karthikeyan et al. [70] proposed advanced security frameworks of Intelligent Transportation Systems (ITS) by integrating Q-learning with DRL methodologies to effectively manage and mitigate DDoS flooding attacks in IoTdynamic vehicular networks. This integration especially targets the complex network environments of Vehicle-to-Infrastructure (V2I) and Vehicle-to-Vehicle (V2V) communications in IoTITS. The adaptability of Q-learning enables it to effectively respond to low-rate and high-rate DDoS attacks, thus significantly enhancing the resilience of Roadside Units (RSUs) and overall network traffic management. This approach proactively adjusts defense strategies in real-time, aligning them with evolving attack patterns to safeguard critical transportation infrastructure. The dynamic adjustment is facilitated by continuously updating the reward mechanisms and policy networks based on real-time traffic and threat analyses, allowing immediate and effective responses to changing threat dynamics.

Frantzy Mesadieu et al. [71] innovatively applied a DRL framework that integrates a DQN for enhanced IDS in IoTSCADA systems. This pioneering approach exploits the synergies between Q-learning and Deep Learning (DL) through what is designated as the ’Q-network.’ The ’Q-network’ is a specialized neural architecture that augments Q-learning with the capacity to analyze and learn from complex data patterns in network traffic, patterns that are often overlooked by conventional detection methods. Conventional IDS, primarily signature-based and anomaly-based systems, frequently struggle against novel and sophisticated cyber threats due to their inherently static nature. While these traditional methods are effective against known threats, they lack the flexibility to adapt to new, evolving challenges. In stark contrast, the Q-network dynamically processes and updates from ongoing data streams, markedly enhancing its ability to identify anomalies that indicate potential security breaches in real-time. The deployment of the Q-network represents a substantial shift away from traditional methodologies by equipping the system with a continuous learning mechanism. Moreover, this mechanism utilizes historical data and continuously adapts to new threat patterns. This crucial adaptive capability significantly contributes to the system’s outstanding detection accuracy of 99.36%.

Hafiz Husnain Raza Sherazi et al. [72] developed a sophisticated hybrid detection system to combat DDoS attacks in IoV networks, integrating fuzzy logic and Q-learning to enhance security measures. Fuzzy logic effectively processes ambiguous and uncertain data, allowing dynamic responses to diverse attack patterns. Q-learning adapts defensive strategies based on real-time attack data, improving the system’s reactivity and adaptiveness. The simulations demonstrate significant improvements in buffer size efficiency, energy consumption, response time, and network throughput, underscoring the system’s capability to maintain operational performance even under attack conditions.

### 4.6. Multi-Armed Bandit (MAB)

MAGPIE represents a significant advancement in IDS for smart homes, developed by Ryan Heartfield et al. [73]. This system dynamically adjusts its anomaly classification decisions using a non-stationary-MAB approach and updates its probabilistic cluster-based reward functions based on silhouette scores. These adjustments are crucial as they account for the non-stationary behaviors typical of smart home environments and the evolving interactions of users with their devices. In this context, "valid" refers to the system’s ability to operate effectively under realistic conditions. MAGPIE demonstrated high efficiency and accuracy, especially in recognizing novel cyber-physical threats and adapting to new user behavior patterns. Experimental evaluations show that MAGPIE achieves an anomaly detection accuracy rate of up to 93%, underscoring its potential to enhance security in smart homes significantly.

Radoglou-Grammatikis et al. [74] presented a cutting-edge Intrusion Detection and Prevention System (IDPS) designed specifically for industrial healthcare systems. By integrating SDN with RL, this study addressed the critical vulnerabilities inherent in the IEC 60 870-5-104 protocol, a widely used standard in such systems. The DRL component is especially noteworthy, as it models the mitigation strategy as a stationary multi-armed bandit problem, which is optimally solved using Thompson sampling. The detection accuracy and F1 score achieved by the IDPS are 0.831 and 0.8258, respectively, indicating high precision in identifying threats. Additionally, the mitigation accuracy reaches 0.923, showcasing the system’s robustness in neutralizing potential cyberattacks.

### 4.7. Deep Deterministic Policy Gradient (DDPG)

Chengming Hu et al. [75] proposed a novel IDS for the smart grid, named RL-Based Adaptive Feature Boosting (AFB). This system leverages multiple AutoEncoders to extract critical features from the multi-sourced, heterogeneous data generated in smart grid environments. These features are then utilized to train a Random Forest (RF) classifier. The RL-AFB system employs the DDPG algorithm to dynamically adjust the feature sampling probabilities based on their contribution to classification accuracy. The application of DDPG significantly improved classification performance, achieving a 97.28% accuracy on the Hardware-In-the-Loop (HIL) security dataset and outperforming other methods on the WUSTIL-IIOT-2021 dataset.

Laisen Nie et al. [76] proposed a novel IDS explicitly designed for green IoT systems, utilizing DDPG-based DRL. This IDS addresses the significant challenges associated with traditional methods, such as slow detection speeds and high false alarm rates. By implementing DDPG, the system efficiently predicts and schedules network traffic flows, enabling the rapid and accurate detection of DDoS attacks. In the evaluation, the DDPG-based IDS substantially improved detection accuracy and speed. The system achieved a True Positive Rate (TPR) of 99.98% and 100% in two different datasets, significantly outperforming conventional methods such as SRMF and MWM, which showed TPRs of 92.48% and 82.76%, respectively. Additionally, the system maintained a low False Positive Rate (FPR), indicating its ability to minimize false alarms while adapting to dynamic network conditions. The dynamic threshold adjustment capability of DDPG was especially effective, reducing the Time Relative Error (TRE) to 0.2752 and 0.3721 in two datasets, compared to 0.9838 and 0.9964 for traditional methods. This result highlighted the system's ability to adapt in real-time to fluctuating network conditions, thus significantly enhancing the real-time security of IoT networks.

Noora Mohammed et al. [77] proposed a novel approach to enhance the security of FL in IoT systems by utilizing a DDPG-based reputation management mechanism. This method addresses the dynamic nature of worker behavior in FL by continuously optimizing reputation thresholds, critical in identifying and mitigating the risks posed by unreliable or malicious participants. Unlike traditional approaches such as the static reputation threshold models, which set a fixed threshold that may not adapt well to varying worker behaviors, or FedAvg, which lacks any built-in mechanism for evaluating the trustworthiness of workers, the DDPG-based approach allows for more nuanced and effective decision-making. Additionally, the study compares the DDPG method with DQN-based reputation models, highlighting that while DQN can handle discrete action spaces, it struggles with the continuous and high-dimensional action spaces often encountered in FL settings. In contrast, DDPG excels in such scenarios, leading to a more robust selection of reliable workers and ultimately improving the overall accuracy and stability of the FL model. The results presented in the study show that the DDPG-based system improves model accuracy by over 30% compared to conventional methods, including FedAvg, DQN-based reputation models, and static reputation threshold methods.

### 4.8. Inverse Reinforcement Learning (IRL)

Juan Parras et al. [78] proposed using Inverse Reinforcement Learning (IRL) to enhance defense mechanisms against intelligent backoff attacks in wireless networks. They introduced two IRL-based strategies that operate under partial observability, detecting attacks by analyzing deviations from normal behavior. Unlike traditional methods, these mechanisms generalize from known attacks to predict and counteract previously unseen strategies, significantly improving network resilience. While IRL is distinct from DRL, which focuses on inferring reward functions rather than optimizing policies, it can be augmented with DL for handling complex scenarios. This study's contribution is crucial as it provides a flexible, adaptive defense framework that requires minimal assumptions about attack types, addressing the challenge of increasingly sophisticated cyber threats.

### 4.9. Proximal Policy Optimization (PPO2)

Sumegh Tharewal et al. [79] proposed IDS is sourced from the U.S. Department of Energy's Oak Ridge National Laboratory. This dataset is designed explicitly for IIoT systems, encompassing various types of network traffic data representative of real-world IIoT scenarios. The dataset includes standard traffic data and a wide range of malicious attack data, ensuring a robust testing ground for the IDS. Specifically, the dataset comprises 26 distinct features, each representing different aspects of network traffic, such as packet size, protocol types, and time-based features. These features are crucial for identifying patterns and anomalies in IoT networks, which indicate normal operations or potential security breaches. The attack types covered in the dataset include DoS, reconnaissance attacks, malicious command injections, and other forms of cyber threats that are common in IIoT systems. In the study, LightGBM filters out the least important features, reducing the feature set to those most relevant for IDS. This process ensures that the IDS operates efficiently without compromising on detection accuracy. The dataset's diversity and its alignment with real-world IIoT conditions make it an ideal choice for training and validating the IDS, allowing for the demonstration of the system's ability to detect and respond to various types of network threats with high precision and speed. The rigorous testing on this dataset highlights the IDS's superior performance, achieving a detection accuracy of 99.09% and demonstrating its practical applicability in securing IIoT systems.

### 4.10. One-Shot Learning Techniques

Nouf Saeed Alotaibi et al. [80] introduced a dynamic IDS tailored for smart city systems, utilizing RL techniques, specifically one-shot learning, enhanced with DRL capabilities. This approach is specially crafted to tackle the inherent dynamic challenges associated with multi-access edge computing architectures, e.g., managing the variability in network behavior and the potential for increased security vulnerabilities. The model efficiently mitigates zero-day attacks and other emergent threats in IoT-based networks. One-shot learning enables the system to adapt to novel scenarios rapidly using minimal training data. In contrast, the DRL component effectively uses sparse experiential data to make informed real-time decisions. This dual functionality significantly boosts the system's resilience, maintaining robust security against the evolving landscape of network threats. Empirical results indicate a detection accuracy rate of approximately 98.8% with low false positive rates, underscoring the model's effectiveness in real-world applications.

### 4.11. Double Deep Q-Network (DDQN)

Shi Dong et al. [81] combines the strengths of supervised and unsupervised learning to detect anomalous traffic in network data effectively. A key element of this approach is the DDQN, a variant of DRL that enhances the precision and robustness of network anomaly detection. In this method, an autoencoder is first used to reconstruct network traffic features, fed into a deep neural network integrated into the DDQN framework for classification in IoT. This layered approach allows the model to dynamically adjust to varying traffic patterns and accurately identify known and unknown network anomalies. The implementation of K-Means clustering further aids in detecting unknown attacks by categorizing traffic features into distinct clusters without prior labeling, thus reducing the reliance on manually labeled data. The improvements brought about by this semi-supervised method are significant, especially when compared to traditional ML models (e.g., DT, SVM, and RF). The SSDDQN method achieved an accuracy of 83% in binary classification and improved the F1-Score by approximately 6% over the traditional DDQN model. In five-class classification tasks, the SSDDQN model reached an F1-Score of 79.43% and demonstrated a 5% improvement in detection rate (DR) compared to DDQN. These numerical results underscore the model's capability to reduce time complexity while enhancing critical performance metrics such as accuracy, F1-Score, and detection rate, making it especially effective in real-time network environments.

Mohammad Alauthman et al. [82] presented a sophisticated approach to detecting peer-to-peer (P2P) botnet traffic, which is difficult to identify due to its decentralized and evasive characteristics. The study developed a system that integrates a refined traffic reduction mechanism with RL techniques, specifically targeting the improvement of botnet detection accuracy. The proposed approach was thoroughly tested on real-world network traffic, yielding a high detection accuracy of 98.3% and maintaining a low false positive rate of 0.012%. This combination of network traffic analysis with an adaptive RL framework marks a significant advancement in the domain, as it effectively identifies bot-related activities with minimal disruption to network performance. The RL-based method employed is carefully tailored to the unique challenges posed by P2P botnets, offering a robust and highly adaptive solution.

Praveena et al. [83] introduced an optimal DRL approach specifically designed for an IDS in Unmanned Aerial Vehicles (UAVs). The approach leverages DDQN combined with Black Widow Optimization (BWO) to enhance the IDS's performance. By utilizing BWO for hyperparameter tuning, the model fine-tunes the network parameters, significantly improving the IDS's capability to accurately predict and prevent threats in UAV networks. This advanced method strengthens the learning process through DDQN. It optimizes the detection process to handle the complexities of high-dimensional data and intricate attack vectors commonly encountered in UAV networks. The efficacy of this IDS was tested using the NSL-KDD dataset, where it achieved remarkable performance metrics: a precision of 0.985, recall of 0.993, F-measure of 0.988, and accuracy of 0.989. These results affirm the model's effectiveness in real-world scenarios, demonstrating its potential as a robust IDS solution for UAV networks. Integrating DDQN with BWO deepens the IDS's learning capability and optimizes its detection process, making it highly effective in addressing the unique security challenges posed by UAV networks.

### 4.12. Policy Gradient-DRL (PG-DRL)

Vikas Juneja et al. [84] tackled the problem of energy-draining vampire attacks in WSNs, which disrupt network efficiency by extending data transmission routes, leading to rapid energy depletion in sensor nodes. The authors introduce a dual-strategy approach combining a cooperative trust calculation mechanism with a PG-DRL algorithm, demonstrating quantifiable improvements in network performance. The cooperative trust mechanism plays a pivotal role in detecting malicious nodes by analyzing deviations in energy consumption patterns. The cooperative trust matrix, generated by leveraging the residual energy of connected nodes, enables the detection of vampire nodes with an accuracy rate that consistently reaches 100% in controlled scenarios. This method is especially effective in networks with varying environmental conditions, where traditional threshold-based methods might fail. Simultaneously, the PG-DRL algorithm optimizes routing paths by dynamically selecting the next-hop node, ensuring minimal energy consumption and maintaining high levels of trust in the IoT network. The PG-DRL approach significantly reduces unnecessary energy expenditure, with simulations indicating a 3% increase in network lifetime compared to the benchmark Dynamic Source Routing (DSR) protocol. Moreover, the methodology also improves the detection ratio by 20% compared to existing state-of-the-art schemes.

### 4.13. State-Action-Reward-State-Action (SARSA)

Abbasgholi Pashaei et al. [48] developed cutting-edge IDS for industrial control networks, focusing on detecting and preventing critical threats such as Man-in-the-Middle (MITM) and DDoS attacks. The system utilizes the SARSA algorithm, a model-free, on-policy RL method. Unlike DRL, SARSA updates its policy based on current action dynamic adaptability in real-time network environments. The IDS is structured with a dual-agent model: one agent simulates the network environment to challenge the classification agent, which enhances its ability to detect complex attacks. This adversarial setup, combined with a honeypot that mimics realistic network behaviors, allows the system to effectively lure and learn from potential attackers. The system demonstrated over 99% accuracy and an F-measure of 0.98 in extensive tests, outperforming traditional IDS models.

### 4.14. Grasshopper Optimization Algorithm (GOA)

Alok Kumar Shukla et al. [45] presented an innovative approach to enhancing IDS through advanced optimization and RL techniques. The study introduced a novel variant of the Grasshopper Optimization Algorithm (GOA), termed the opposition self-adaptive GOA, which incorporates mutation and perceptive strategies to improve the selection of relevant features for anomaly detection. To further bolster the system's detection capabilities, the study integrated the RL method, specifically the Gain Actor-Critic (GAC) approach, into the SVM. This hybrid model was evaluated using standard IDS datasets, including NSL-KDD, AWID, and CICIDS 2017. The results showcased the model's superior performance, achieving a detection accuracy of 99.71% on NSL-KDD, 99.11% on AWID, and 99.61% on CIC-IDS 2017, along with impressively low false-positive rates of 0.009, 0.091, and 0.052, respectively.

### 4.15. Multiagent Deep Deterministic Policy Gradient (MADDPG)

Delali Kwasi Dake et al. [46] introduced a DRL framework for SDN and IoT, focusing on routing optimization and DDoS detection/prevention using the MADDPG algorithm, an extension of DDPG for multi-agent environments. This framework allows multiple RL agents to work cooperatively or competitively to enhance network performance metrics, e.g., delay, jitter, and packet loss rate. These improvements contribute to better IDS capabilities by ensuring stable network conditions, which reduce false positives and negatives, thus allowing more accurate detection of genuine threats. Extensive simulations show that the MADDPG-based framework outperforms traditional DDPG, especially in complex, dynamic network scenarios, highlighting its effectiveness in improving security and performance in SDN-IoT systems.

### 4.16. Dueling Double Deep Q-Network (D3QN)

Xuecai Feng et al. [44] developed the Security Defense Strategy Algorithm (SDSA), a novel approach to enhancing IoT security using DRL. The SDSA leverages the D3QN, which integrates the strengths of Dueling and Double Q-learning methodologies, to optimize the allocation and coordination of defense resources. By simulating adversarial security scenarios involving multiple defenders and attackers, the SDSA enables the adaptive learning of strategic behaviors that enhance defense effectiveness. The efficacy of the SDSA is demonstrated through empirical data, where the method showed an improvement in defense effectiveness by approximately 87% and 85% compared to the MADDPG and OptGradFP models, respectively, in scenarios involving a single attacker. In more complex scenarios with multiple attackers, the SDSA outperformed the MADDPG and OptGradFP models by 65% and 60%, respectively.

### 4.17. Neural Fitted Q-Iteration (NFQ)

Sengathir Janakiraman et al. [47] presented a sophisticated approach to combating DDoS attacks in cloud environments supported by fog computing. The study introduced a novel DDoS mitigation scheme that leverages the integration of DRL and LSTM networks to enhance the detection and mitigation processes. The scheme employs the NFQ algorithm with LSTM to dynamically analyze and respond to malicious traffic. This combination effectively manages high-dimensional state spaces and temporal dependencies, improving the system's ability to classify and filter out infected packets while ensuring service availability. The framework also incorporates SDN technology in IoT, the fog layer, enabling a robust defense mechanism against DDoS attacks by efficiently managing and controlling network traffic across cloud and fog environments. The experimental results demonstrate the proposed scheme's effectiveness, with the RDL-LSTM-2.2 variant achieving a classification accuracy of 98.34%, precision of 98.88%, and recall of 98.76%. Additionally, the proposed scheme reduced packet latency by 10.42% and computational complexity by 12.96% compared to existing approaches. The scheme also achieved a 99.59% training accuracy and a 98.98% testing accuracy, demonstrating its superior performance in mitigating DDoS attacks in a fog-assisted cloud environment.

Table 1 compares thirty-five state-of-the-art DRL-based IDS applied to IoT. It highlights each technique's advantages, limitations, and specific use cases, offering a concise overview for researchers and practitioners.

Table 1: State-of-the-Art DRL-based IDS for IoT Security: A Comparative Analysis

| Paper | Advantages | Limitations | Use Case/Application |
|---|---|---|---|
| [55] | High accuracy (96–97%) in single- and multi-layered attacks, low false positive rate. | Focused primarily on IP blocking, limited exploration of other attack types. | Dynamic IP blocking for IoT systems with scalability for Metaverse applications. |
| [56] | Robust design for adversarial attack evaluation. | Limited implementation in real-world IIoT scenarios. | Identifying vulnerabilities in IIoT systems during training and operation. |
| [57] | Dual-layer edge and cloud security with high accuracy (98.2%). | Dataset limitations; results may not generalize across diverse IoT networks. | Multi-attack classification for UNSW-NB-15 and IoTID20 datasets. |
| [58] | Excellent performance on ISOT-CID and NSL-KDD datasets, achieving up to 99.66% accuracy. | May require high computational resources due to federated learning. | Federated IDS with privacy-preserving mechanisms. |
| [59] | Effective decision-making under adversarial conditions, with a 94.6% win rate. | Focused on fog environments; applicability to other contexts unverified. | Intrusion response in fog computing environments using Minimax-DQN. |
| [60] | Enhanced IoT edge device defenses with k-NN and DTW integration. | Limited scalability for large-scale IoT networks. | Edge device security using zero-sum game frameworks. |
| [61] | Early and late malware detection phases achieving up to 99.8% accuracy. | Complex implementation with higher training overhead. | Malware botnet detection in IoT networks using adaptive strategies. |

| Paper | Advantages | Limitations | Use Case/Application |
|---|---|---|---|
| [62] | Improved detection speed and accuracy compared to traditional ML methods. | Limited exploration of real-time environments. | Simulation-based evaluation using NSL-KDD and AWID datasets. |
| [63] | High anomaly detection accuracy (98.7%) with minimal false alarms. | Focused on smart cities, with limited generalizability to other IoT domains. | Security enhancement in smart cities through MDRL. |
| [64] | Faster rerouting and attack mitigation in SDN-based IoT systems. | Focused only on Link Flooding Attacks. | Mitigation of LFAs in IoT networks using Rényi entropy and DQL. |
| [65] | Strong mitigation of routing threats in OppIoT networks with 47.9% delivery probability. | May not generalize to more dynamic network environments. | Opportunistic IoT threat detection and routing optimization. |
| [67] | High accuracy in feature selection and threat detection with 96.8% and 99.1%. | High computational overhead during training. | IDS enhancement using graph-based feature optimization. |
| [68] | Effective dual-level threat detection at packet and flow levels with 98.78% accuracy. | Evaluation is restricted to CICDDoS2019, which does not encompass the full diversity of IoT attack scenarios. | Multi-level traffic analysis in IoT networks. |
| [69] | Dynamic adaptation to varying network conditions with over 95% detection accuracy. | Limited evaluation in diverse IoT deployment scenarios. | Collaborative DDoS detection for IoT edge devices. |
| [70] | Adaptive defenses against DDoS attacks in vehicular networks. | Focused only on V2I and V2V communications. | Intelligent Transportation Systems (ITS) security. |
| [71] | Dynamic learning of network traffic patterns, achieving 99.36% accuracy. | Limited scalability for large networks. | SCADA systems and industrial IoT. |
| [72] | Hybrid approach improves detection efficiency and throughput in IoV. | High resource consumption in fuzzy logic processing. | IoV DDoS detection and response systems. |
| [73] | Dynamic classification for non-stationary IoT environments, 93% accuracy. | Focused only on smart home IoT. | Anomaly detection in smart homes. |
| [74] | High mitigation accuracy (0.923) using multi-armed bandit modeling. | Limited applicability beyond industrial healthcare systems. | Intrusion detection in industrial healthcare. |
| [75] | Improved feature boosting with 97.28% accuracy on HIL datasets. | Limited testing on non-smart grid environments. | Smart grid security through adaptive feature optimization. |
| [76] | High accuracy (99.98%) in detecting DDoS attacks in green IoT systems. | Focused solely on energy-efficient networks. | Sustainable IoT network protection. |
| [77] | Effective reputation management in federated IoT systems with over 30% accuracy improvement. | Focused only on reputation-based FL models. | Enhancing FL security in IoT. |
| [78] | Generalizes to predict unseen attack strategies, improving resilience. | Focused on backoff attacks only. | Wireless network security under partial observability. |
| [79] | Superior detection accuracy (99.09%) using diverse IIoT datasets. | Focused only on lightGBM filtering for IIoT. | Enhanced IDS for IIoT networks. |
| [80] | One-shot learning enables fast adaptation to zero-day attacks with 98.8% accuracy. | Complex implementation for sparse data. | Smart city IDS leveraging DRL and one-shot learning. |
| [81] | High F1-Score (79.43%) in five-class classification tasks. | Limited scalability for larger datasets. | Network anomaly detection using semi-supervised DRL. |
| [82] | High accuracy (98.3%) in P2P botnet detection. | Limited evaluation on non-P2P traffic, which restricts its generalizability to other types of botnet architectures. | Identifying decentralized botnet activities. |
| [83] | Advanced detection metrics with over 98% precision and recall. | Limited to UAV IDS, potentially requiring adaptation for broader IoT applications. | Unmanned Aerial Vehicle network protection. |
| [84] | 100% detection rate for vampire attacks with energy-efficient routing. | Limited exploration of non-energy threats. | Secure routing in WSNs against vampire attacks. |
| [48] | High accuracy (99%) and dynamic adaptability in real-time. | High resource overhead for dual-agent design. | Industrial network security against MITM and DDoS. |
| [45] | Excellent accuracy (99.71%) across multiple datasets. | Focused on GOA optimization only. | Multi-feature IDS optimization in IoT. |
| [46] | Superior network performance using MADDPG for dynamic scenarios. | Limited real-world implementation. | Routing and DDoS mitigation in SDN-IoT. |

| Paper | Advantages | Limitations | Use Case/Application |
|---|---|---|---|
| [44] | Enhanced defense coordination with over 87% effectiveness improvement. | High training complexity for multi-agent setups. | Multi-agent security defense in IoT. |
| [47] | Low latency and high accuracy (98.98%) in cloud-fog environments. | Focused on fog-supported IoT only. | Fog-assisted DDoS detection in cloud environments. |

## 4.18. Distribution of Models in Literatures

Figure 8 provides a comprehensive view of the distribution of various DRL-based models utilized in IDS in IoT systems. This distribution highlights the prevalence and application-specific suitability of different models, reflecting both their strengths and the unique challenges they address in IoT contexts.

The DQN dominates with a 25.7% share, underscoring its significant role in IoT security. DQN's prominence is attributable to its robustness in handling large state and action spaces, typical in IoT systems characterized by numerous interconnected devices and complex interactions. The DQN model leverages experience replay and off-policy learning, allowing it to store past experiences and learn from them effectively. This capability is crucial in IoT scenarios requiring real-time decision-making and adaptability to dynamic conditions. The model's ability to optimize policies by evaluating the Q-values of actions in given states makes it especially effective in environments where exploration and exploitation must be balanced.

The MDRL model accounts for 11.4% of the usage, highlighting its adaptability in tailored IoT applications. MDRL models often involve modifications to standard DRL architectures to handle better IoT systems' specific constraints and requirements, such as limited computational resources, energy efficiency, and rapid response to threats. The modifications typically aim to enhance the model's ability to operate in highly variable and unpredictable environments, which is common in IoT networks.

DQL and SAC each represent 8.6% of the applications, indicating their critical roles in specific IoT scenarios.

DQL is an extension of Q-learning that integrates DL to manage continuous action spaces more effectively. This makes DQL highly suitable for IoT systems where decisions are not binary but fall on a spectrum of possible actions, such as adjusting the intensity of an IoT device's response to an external threat. DQL's application in high-dimensional sensory input processing, especially in sensor networks and robotics in IoT, ensures that systems can respond more precisely and effectively to nuanced threats.

SAC is especially favored for its capacity to maintain stability and reliability in learning processes even under dynamically changing policy conditions. This attribute is essential in IoT applications that require long-term strategic planning (e.g., smart grids or autonomous systems), where the environment and network conditions can vary significantly over time. SAC's use of stochastic policies, combined with entropy regularization, enables it to balance exploration and exploitation more effectively, making it robust against overfitting specific scenarios and ensuring broader applicability across varying IoT contexts.

Other models, each contributing 2.9% to the distribution, are chosen for specialized applications, reflecting their tailored utility in specific IoT systems:

PG-DRL and DDPG are especially effective in configurations with continuous action spaces. They are employed in scenarios requiring high-dimensional decision-making processes, where traditional discrete action models like DQN might struggle. PG-DRL, for instance, directly optimizes the policy by following the gradient of expected rewards, making it suitable for complex tasks where the policy space is large and continuous. DDPG extends this by incorporating deterministic policies, which improve efficiency in environments where exploration noise needs to be minimized, such as in highly regulated or safety-critical IoT systems.

Q-learning and SARSA are valued for their simplicity and convergence properties, making them reliable for applications where the action space is discrete and well-defined. Q-learning is especially robust in environments where the agent needs to learn from an off-policy perspective, adapting to environmental changes even if they are not directly related to the actions taken. SARSA, being an on-policy algorithm, is more suitable in environments where the policy being evaluated is the one being improved, providing a more stable learning process in such scenarios.

IRL is employed when the objective is to infer the underlying reward structure from observed behavior rather than optimizing a predefined reward. This is especially useful in surveillance and anomaly detection systems in IoT, where the goal is to understand and replicate the decision-making processes of expert systems or humans. IRL's ability to generalize from observed data makes it a powerful tool for environments where the reward structure is not explicitly known or is difficult to define.

One-shot learning techniques are critical when data is scarce or the system needs to adapt rapidly to new and unseen scenarios. In IoT systems, where gathering large amounts of labeled data can be challenging due to devices' distributed and heterogeneous nature, one-shot learning allows systems to generalize from minimal examples, ensuring quicker adaptation and response times.

Stationary and MAB models are utilized in scenarios where the reward distributions change over time, which is common in IoT systems subject to fluctuating network conditions and varying threat levels. These models are instrumental in dynamic resource allocation problems, where the system must continuously adapt its strategy to maximize the long-term rewards.

DDQN and D3QN improve upon standard DQN by addressing the overestimation bias inherent in Q-learning, which can lead to suboptimal policies. DDQN uses separate networks to select and evaluate actions, reducing this bias. At the same time, D3QN further refines this by decoupling the value and advantage functions, allowing the model to better distinguish between valuable actions in similar states. These improvements make them highly effective in complex IoT systems where fine-tuned action selection is critical.

MADDPG and NFQ are applied in more complex IoT systems that involve multiple agents or require a nuanced understanding of time-sequential data. MADDPG is especially useful in environments where multiple IoT devices must collaborate or compete, such as in networked control systems or cooperative sensor networks. At the same time, NFQ excels in scenarios where the system must learn from sequences of decisions, making it ideal for temporal analysis in IoT security.

The diversity of models employed in DRL-based IDS for IoT reflects the multifaceted challenges these systems face, from handling continuous and high-dimensional action spaces to ensuring real-time responsiveness and adapting to evolving threats. The unique demands of the application often dictate the selection of a specific model, the nature of the IoT environment, and the specific objectives of the IDS, whether that is optimizing resource usage, minimizing response times, or improving detection accuracy.

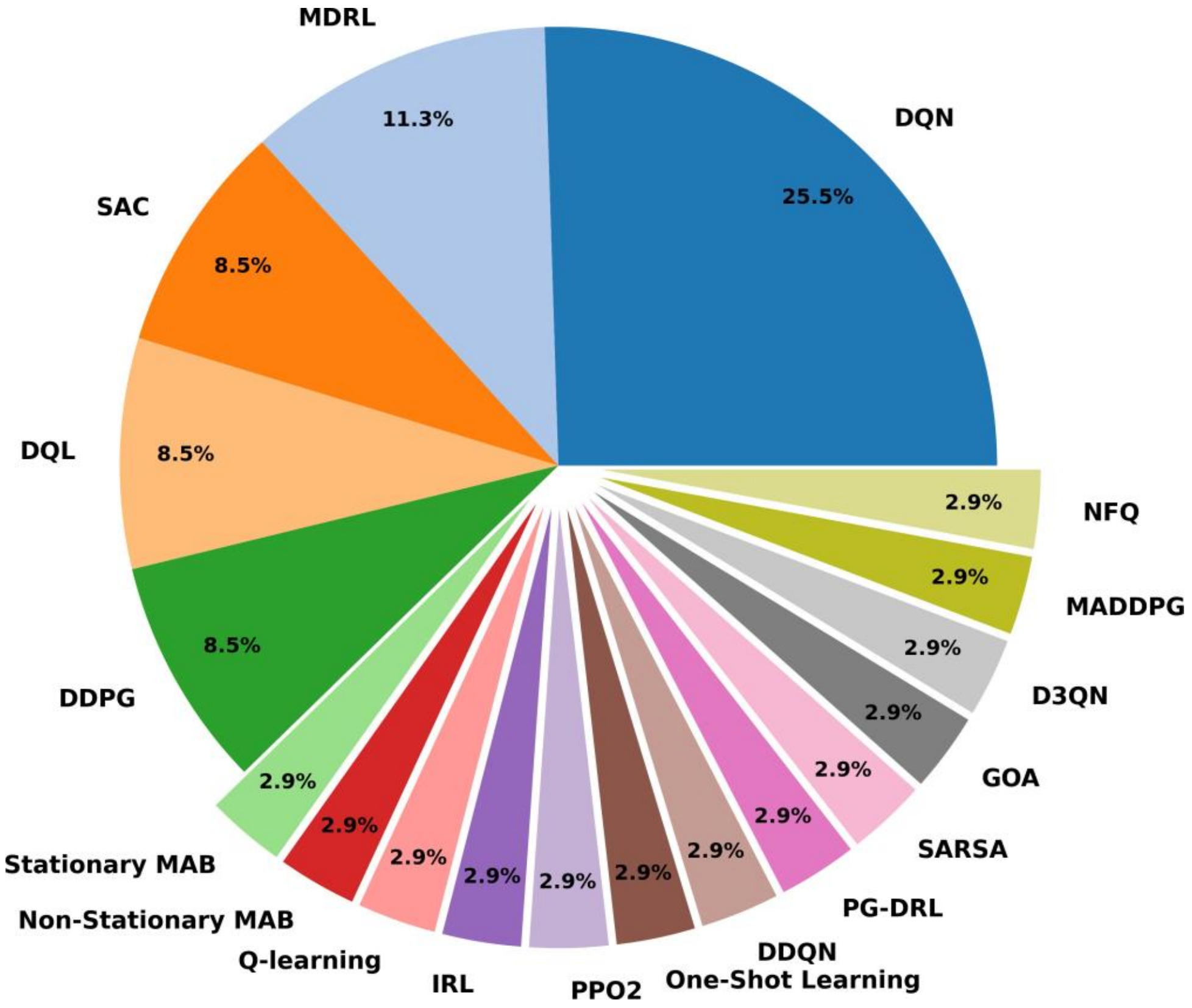

**Figure 8:** Usage Frequency of Different Models in DRL-based IDS in IoT.

**Table 2**
DRL-based IDS applications in various IoT contexts.

| IoT Sector | Paper Reference | Brief Analysis |
|---|---|---|
| Smart Home Security | [55], [63], [64], [65], [56] | Focuses on enhancing security measures in smart homes through adaptive threat detection and real-time response to diverse threats like DDoS and cyber-attacks. |
| Industrial IoT (IIoT) | [69], [66], [57], [70], [47], [71], [85] | Discusses complex decision-making and operational optimization in industrial systems, addressing high-stakes security challenges. |
| Healthcare IoT Systems | [74], [75], [78], [79] | Applies DRL to ensure critical accuracy and privacy, adapting to the sensitive nature of healthcare data and operations. |
| Transportation Systems | [70], [82], [83], [62], [60], [77] | Enhances security in transportation by managing dynamic environments and securing high-speed vehicular networks against sophisticated attacks. |
| Smart Cities | [58], [80], [68], [76], [81], [59], [44], [46] | Utilizes DRL to manage complex urban IoT systems, optimizing security across dense city networks and enhancing city-wide communication protocols. |
| General IoT Security | [67], [84], [48], [45], [61] | Explores the broader application of DRL across various IoT domains, emphasizing adaptive learning and security enhancements in diverse operational contexts. |

## 4.19. The existing DRL-based IDS

The application of DRL-based IDS spans various IoT contexts, each addressing distinct security challenges, as summarized in Table 2. In smart homes, DRL's adaptive threat detection and real-time response capabilities are vital for securing complex networks of diverse devices. In IIoT, DRL-based IDS protects critical infrastructure by optimizing decision-making and mitigating sophisticated threats that could disrupt operations. In healthcare IoT, DRL ensures accuracy in threat detection and protecting sensitive patient data, a vital necessity given the high stakes involved. Transportation systems, especially vehicular networks, benefit from DRL's ability to secure dynamic environments against advanced cyberattacks. Smart cities, with their intricate IoT ecosystems, leverage DRL to safeguard essential services and maintain operational efficiency. The versatility and adaptive learning capabilities of DRL make it especially beneficial in IoT applications, where the ability to learn from evolving threats and improve detection accuracy continuously is critical. This positions DRL as a cornerstone technology for enhancing security in the increasingly complex landscape of IoT, ensuring robust protection across diverse sectors.

**Table 3**
Overview of types of attacks covered by each dataset

| Dataset Name | Types of Attack Covered |
|---|---|
| ISCX 2012 IDS | Brute Force SSH, Brute Force FTP, Infiltration, HTTP DoS, DDoS, Botnet |
| Microsoft Malware | Malware (Viruses, Worms, Trojans, Spyware, Adware, etc.) |
| N-BaIoT | DDoS, DoS, Reconnaissance, MITM |
| UNSW-NB-15 | Fuzzers, Analysis, Backdoors, DoS, Exploits, Generic, Reconnaissance, Shellcode, Worms |
| CICIDS2017 | DoS, DDoS, Heartbleed, Bot, Infiltration, Web Attacks, PortScan, FTP-Patator, SSH-Patator, etc. |
| IoTID20 | Backdoor, Analysis, Fuzzing, DoS, Exploits, Generic, Reconnaissance, Shellcode, Worms |
| NSL-KDD | DoS, Probe, U2R, R2L, etc. |
| AWID | Flooding, Injection, Impersonation, Miscellaneous |
| CSE-CIC-IDS2018 | Brute Force, DoS, DDoS, SQL Injection, Heartbleed, etc. |
| WUSTL-IIoT-2021 | APTs, MITM, Side-channel Attacks, Zero-day Exploits, etc. |
| WUSTL-IIoT-2018 | Advanced Persistent Threats, Evasion Techniques, Side-channel Attacks |
| MedBIoT | Early-stage malware activities (e.g., scanning, reconnaissance) |
| TON-IoT Dataset | Botnet, Command Injection, Malware Propagation |
| DS20S Trace | Robust Feature Extraction in Noisy and Heterogeneous IoT Systems |
| Hogzilla | CTU-13 Botnet, Illegitimate Packets |
| SCADA-based Power System | The famous Stuxnet worm that damaged nuclear machinery in Iran. |
| CIC-DDOS2019 | DDoS Attacks (e.g., UDP Flood, HTTP Flood, SYN Flood) |
| BoT-IoT | DDoS, DoS, OS and Service Scan, Keylogging and Data exfiltration attacks. |
| ISOT-CID | Various IoT-specific attacks and vulnerabilities |
| BoTNeTIoT-L01 | Automated IoT botnet attack scenarios, traffic anomalies |
| IEC 60 870-5-104 | Protocol-specific attacks targeting energy sector communications |

## 5. The datasets are used for DRL-based IDS

To address RQ2, we thoroughly examine the datasets used in DRL-based IDS for IoT, detailed in Table 3. We classify these datasets to show their specific role and impact on IoT security, pointing out how each contributes to the enhancement of the IDS capability in IoT.

#### 5.0.1. *Real-World Network Trafic Simulation*

Datasets such as ISCXIDS2012 [86], UNSW-NB 15 [87], and CICIDS2017 [88] are crucial for simulating realistic network environments for IDS training and evaluation. These datasets include labeled data representing various attack scenarios commonly observed in real-world networks, such as DDoS, brute force attacks, and infiltration attempts.

#### 5.0.2. *Specialized IoT systems Attacks*

Specialized datasets such as WUSTL-IIoT-2018 [89], WUSTL-IIoT-2021 [90], and N-BaIoT [91] focus specifically on industrial and home IoT systems, offering scenarios including advanced persistent threats (APTs) and zero-day exploits. For example, the WUSTL-IIoT datasets capture traffic data from industrial IoT devices under various attack scenarios, including APTs, which are long-term targeted attacks aiming to steal data or disrupt operations. These datasets are critical for developing IDS that target the unique vulnerabilities of IoT devices and networks, such as botnet attacks (e.g., Mirai botnet variants), which exploit common IoT weaknesses like default credentials and outdated firmware.

#### 5.0.3. *Emerging and Evolving Threats*

The Hogzilla [92] and IoTID20 [93] datasets are instrumental in reflecting the evolving nature of cyber threats, especially those involving sophisticated malware and botnet configurations targeting IoT devices. For example, Hogzilla includes scenarios simulating advanced malware behaviors, such as polymorphic and metamorphic techniques, which allow malware to change its code to evade detection. IoTID20 captures the traffic patterns associated with emerging threats like cryptojacking and data exfiltration through botnets.

#### 5.0.4. *Protocol-Specific Security*

The IEC 60870-5-104 [94] dataset tests communication protocols used in critical sectors such as the energy industry, especially in SCADA systems. This protocol is widely used for data exchange between control centers and substations. Vulnerabilities in this protocol can allow attackers to manipulate control signals, leading to catastrophic outcomes such as power outages.

#### 5.0.5. *Wireless Network Security*

The AWID dataset [95] focuses on attacks in wireless IoT networks (e.g., flooding, injection, and impersonation). This dataset is especially relevant for training IDS to protect wireless communications in IoT systems, where security risks are heightened due to the inherent vulnerabilities of wireless protocols like Wi-Fi. For instance, the dataset includes scenarios of Wi-Fi flooding attacks, where an attacker overwhelms a network with traffic to cause a denial of service.

#### 5.0.6. Botnet and DDoS Focus

BoT-IoT [96] and CIC-DDoS2019 [97] datasets offer large volume data on botnet and DDoS attacks, predominantly in the disruption of IoT operations. Therefore, these datasets facilitate the establishment of IDS, which can identify and counter the effects of the interruption in case of large-scale coordinated attacks started from multiple compromised devices.

#### 5.0.7. Sector-Specific and Critical Infrastructure

Sector-specific datasets such as TON-IoT [97], MedBIoT [98], and SCADA-based power systems [99] address the unique security challenges in specialized IoT systems. For example, TON-IoT focuses on teleoperated networks and includes data related to malware propagation and command injection attacks, which are common in industrial IoT. MedBIoT is especially important for securing medical IoT devices, where the integrity and confidentiality of patient data are paramount. Also, the SCADA-based power system dataset simulates attacks on critical infrastructure, including scenarios similar to the Stuxnet attack.

### 5.1. Dataset Utilization for Robust DRL-based IDS

IDS have been tested using various datasets (e.g., ISCXIDS2012 [86], UNSW-NB15 [87], and CICIDS2017 [88]). These datasets emulate real-world attack scenarios, including DDoS, brute force, and infiltration attacks, enabling IDS to demonstrate robust performance across diverse operational conditions.

Specialized datasets, such as WUSTL-IIoT, N-BaIoT, and Hogzilla, are especially valuable for enhancing IDS accuracy. They allow IDS to focus on specific threats(e.g., vulnerabilities in industrial IoT or modern malware, enabling finely tuned defenses tailored to these contexts). For instance, using the WUSTL-IIoT dataset, IDS can be trained to detect advanced persistent threats in industrial settings. In contrast, the N-BaIoT dataset helps identify and mitigate botnet activities specific to IoT devices. The proactive integration of new datasets, such as IoTID20, which captures emerging threats such as cryptojacking, and protocol-specific datasets such as IEC 60870-5-104, further strengthens IDS capabilities. By incorporating these datasets, IDS can adapt to evolving threat landscapes and better protect critical infrastructure from specialized attacks. Using sector-specific datasets, such as those for SCADA systems, medical IoT devices, and teleoperated networks, is crucial for refining IDS solutions. These datasets ensure that IDS are generally effective and capable of addressing the unique challenges of specific environments.

## 6. Discussion

By integrating DRL into IDS, these systems gain the capability to dynamically adapt and respond to new and evolving threats, a critical advancement given the complexity and continuous expansion of IoT networks. We present a detailed analysis of states, actions, and rewards across selected works in our SLR to demonstrate the transformative potential of DRL in enhancing IDS for the IoT. An overview of the specific functionalities, state descriptions, actions implemented, and reward mechanisms developed in these studies is presented in Table 4.

DRL empowers IDS with sophisticated decision-making and adaptability, essential for navigating the intricacies of IoT systems. This adaptability is showcased through diverse state definitions, ranging from simple network traffic patterns to complex, multi-dimensional data vectors, as highlighted in studies, e.g., [55] and [63]. Such dynamic adaptability ensures robust protection against a wide array of cyber threats, thus maintaining the integrity and functionality of critical IoT infrastructure.

DRL's flexibility in action selection allows for a broad spectrum of responses tailored to specific network conditions and threats. This is evident from strategies ranging from basic traffic management (e.g., blocking or allowing traffic) to more complex ones such as adjusting classifier parameters [56] and managing traffic [64]. This strategic application of varied actions helps create a more proactive and intelligent security posture, essential for promptly pre-empting breaches and mitigating threats.

The deployment of DRL-based IDS also brings forth challenges, primarily in designing effective reward mechanisms and managing the computational demands of DRL models. Effective reward mechanisms are crucial as they must encourage desirable behaviors that lead to successful threat mitigation, as shown across various studies [61], [71]. However, these mechanisms must also reflect the complex realities of IoT security, where diverse factors such as device heterogeneity, network dynamics, and threat diversity need to be considered without overwhelming IoT system resources [74]. Moreover, the computational intensity of DRL models poses significant challenges for deployment in resource-constrained IoT systems [75], necessitating further research for optimizing these models for practical applications.

In addition, deploying DRL-based IDS in real-world IoT networks presents additional challenges, including computational constraints, energy efficiency, scalability, and sustainability. IoT devices often have limited processing capabilities, making it challenging to implement resource-intensive DRL models [100]. These devices typically operate on constrained power sources, and continuous IDS operations can lead to rapid energy depletion [101]. The scalability of DRL models is also a concern, as IoT networks can consist of thousands of interconnected devices, complicating the deployment of IDS across large-scale environments [101], [102]. Sustainability is another critical factor, as the energy consumption associated with training and deploying DRL models must align with green computing initiatives.

Several strategies have been proposed to address these challenges. Leveraging edge computing and FL can help distribute computational workloads, reducing the burden on individual IoT devices. Implementing energy-efficient DRL techniques, such as model compression and adaptive activation functions, can minimize power consumption [103]. Integrating renewable energy sources into IoT devices can enhance sustainability. Scalable approaches, including distributed and multi-agent DRL frameworks, can effectively manage large networks [104]. Developing lightweight DRL architectures tailored to the specific constraints of IoT devices ensures that IDS deployments are both efficient and sustainable.

## 7. The future research opportunities in DRL-based IDS

This section identifies the current limitations and gaps in DRL-based IDS, especially in IoT systems. By highlighting these challenges, we emphasize the need for future research to develop innovative solutions that enhance the security and effectiveness of IDS in increasingly complex IoT systems.

**Table 4**
Synthesized overview of DRL contributions to IDS in IoT

| Functionality | State Description | Action Implemented | Reward Mechanism |
|---|---|---|---|
| Traffic Management | Metrics such as frequency of packets, port utilization, traffic distribution [64, 69] | Routing decisions, traffic shaping [64, 69] | Maximizes efficiency, minimizes congestion [64] |
| Threat Detection | Network traffic patterns, anomaly scores from classifiers [55, 63] | Classify traffic as normal or malicious, adjust detection parameters [56, 66] | Positive for accurate detection, negative for errors [56] |
| System Control | System status such as valve states, tank levels [65] | Control actions such as opening/closing valves [65] | Rewards based on system performance and safety criteria [65] |
| Security Policy Enforcement | Detected threats, policy compliance levels [67, 80] | Enforcement actions (e.g., block, allow, quarantine) [80] | Rewards for policy adherence, penalties for breaches [67] |
| Adaptive Security | Varying network conditions, attack signatures [48, 84] | Dynamic adaptation of security measures [48] | Rewards for reducing breaches, adapting to new threats [84] |

**Table 5**
Summary of limitations, open challenges, and future work

| Category | Limitations | Open Challenges | Future Work |
|---|---|---|---|
| **Distributed IDS** | Limited focus on distributed models in IoT. | Develop distributed IDS for resource-limited environments. | Explore lightweight, efficient IDS for IoT using edge computing. |
| **IoT Application Domains** | Gaps in coverage of emerging IoT domains. | Expand research to underexplored IoT areas. | Broaden DRL-based IDS for comprehensive security in new domains. |
| **SDN-IoT and DRL** | Limited research on SDN in DRL-based IDS. | Improve SDN-IoT integration for better security decisions. | Develop scalable DRL models optimized for SDN-IoT. |
| **Reward Mechanisms** | Simplistic rewards in DRL models. | Create adaptive rewards that reflect IoT security complexity. | Design detailed rewards prioritizing threats based on impact. |
| **Resource Optimization** | High computational demands of DRL-based IDS. | Balance energy and computational demands without losing accuracy. | Innovate energy-efficient DRL models for IoT constraints. |
| **FL for Privacy** | Privacy concerns with centralized data in DRL-IDS. | Use FL to enhance privacy while improving models collaboratively. | Develop federated DRL models for localized, privacy-focused processing. |
| **Dynamic Threat Detection** | Challenges adapting to sophisticated attacks with traditional DRL. | Enhance DRL to adjust to new threats dynamically. | Use advanced DRL to learn patterns indicating emerging threats. |
| **Real-World Deployment** | Theoretical models not accounting for real IoT constraints. | Address real-world deployment challenges like hardware limits. | Optimize DRL for efficiency in real IoT systems. |
| **Ethical and Legal Considerations** | Potential biases in ML algorithms. | Ensure transparent, fair, and legally compliant IDS operations. | Implement strategies for ethical compliance and oversight. |
| **Topology Awareness** | Using graph models in complex IoT networks. | Leverage network topology for better threat detection. | Apply graph learning to analyze and protect complex IoT networks. |
| **Live Environment Testing** | Lack of performance evaluation in live scenarios. | Conduct real-world testing of DRL-based IDS. | Test IDS in live IoT systems to validate effectiveness. |
| **Security Policy Post-Detection** | Gap in intelligent response after threat detection. | Enhance IDS with autonomous policy implementation. | Design IDS to execute context-aware security responses. |

## 7.1. Dataset

Traditional datasets (e.g., ISCXIDS2012citeunb2012ids, UNSW-NB15citeunsw2015nb15, and CICIDS2017citeunb2017ids) bring valuable insights. However, the rapid evolution of IoT technologies and threat vectors challenges their relevance. There is a pressing need for datasets that evolve alongside these technologies and threats, enabling DRL models to recognize and mitigate novel and sophisticated attacks adeptly. The heterogeneity inherent in IoT systems, spanning diverse devices, protocols, and applications, presents additional layers of complexity. While datasets such as N-BaIoT [91], BoT-IoT [96], and IoTID20 [93] make significant strides toward embracing this diversity, they may fall short of fully encapsulating the intricate dynamics of real-world IoT networks. This shortfall potentially restricts the generalization capabilities of DRL-based IDS. Another pervasive issue in cybersecurity datasets is class imbalance, where instances of normal traffic significantly outnumber attack instances. This imbalance can skew model training, leading to a propensity for overlooking genuine threats. However, the class imbalance is a well-known issue in ML, and there are established solutions, such as oversampling, undersampling, and synthetic data generation techniques such as SMOTE, which can be employed to mitigate this problem and ensure the development of models proficient in accurately identifying attacks without a predisposition towards the majority class.

Integrating real-world data into the training and testing of DRL-based IDS models is crucial for accurately encapsulating the complexities of network traffic and cyber-attack patterns. For instance, the KDDCup 99 [105] dataset offers a blend of diverse network interactions and simulated attacks, setting a benchmark for evaluating the efficacy of IDS solutions. Precise and consistent data labeling is pivotal, especially in the diverse ecosystems of IoT. The accuracy of DRL-based IDS relies heavily on the quality of data labeling. Inaccurate or inconsistent labels can result in models that either flag benign activities as threats due to oversensitivity or fail to detect actual cyber-attacks due to under-sensitivity. Moreover, the considerable computational resources required to process large and complex datasets (e.g., CICDDoS2019 [97]) pose substantial challenges where computational capabilities are limited. Addressing these demands is vital for harnessing the potential of DRL in IDS solutions, necessitating innovative strategies such as model optimization and the adoption of computationally efficient techniques.

Addressing existing datasets' limitations necessitates exploring and including broader data sources. Emerging datasets (e.g., TON-IoT and Edge-IoTset [106]) bring to the fore fresh perspectives and invaluable data for refining IDS mechanisms. These datasets, with their detailed annotations and comprehensive coverage of benign and malicious traffic flows, underscore the importance of a structured approach to tackling IoT security threats. By incorporating such expansive and detailed datasets, the field can move towards developing DRL-based IDS solutions that are theoretically robust and pragmatically effective in the ever-evolving landscape of IoT security.

#### *7.1.1. Enhancing IoT Datasets for Effective IDS*

Current IoT datasets often lack the diversity and dynamism necessary to capture the evolving threat landscape, posing challenges for developing robust IDS. For instance, the CICIoT2023 dataset addresses these limitations by providing a comprehensive collection of IoT network traffic, including various attack scenarios, thereby enhancing the representativeness of training data [107].

Synthetic data generation techniques, such as generative Adversarial Networks (GANs), have been employed to create realistic and varied data samples to address the gaps in dataset diversity [108]. Moreover, simulation tools, e.g., NS3 [109], Cooja [110], and OMNeT++ [111], are increasingly used to generate synthetic network data for IoT networks. These tools enable researchers to emulate IoT scenarios with diverse devices, protocols, and attack vectors, offering controlled environments to enrich datasets. Furthermore, integrating FL frameworks facilitates collaborative model training across distributed IoT devices without compromising privacy, enabling more robust IDS development [112]. These advancements, coupled with simulation and synthetic data generation, empower IDS to adapt effectively to the complexities of modern IoT ecosystems.

### 7.2. Difficult reproducibility of papers

This gap hinders the scientific community's ability to conduct rapid and effective experimental evaluations of proposed solutions, especially in real-world IoT settings, and limits the reproducibility of research fundamental principles for scientific progress and validation of results.

### 7.3. The Predilection for Value-Based DRL Approaches

Our investigation into recent advancements in DRL-based IDS for IoT highlights a significant trend in the predominant use of value-based DRL methodologies, especially DQN frameworks, as seen in studies, e.g., [66] and [79]. The popularity of DQN is primarily due to its simplicity and effectiveness in environments where decisions involve discrete choices (e.g., categorizing network traffic as benign or malicious). This approach aligns well with the structured nature of many IoT systems, making it a favored choice in IoT-IDS research. However, our analysis reveals a gap in exploring more advanced DRL techniques, especially actor-critic and policy learning frameworks. While DQN is well-suited for scenarios with discrete actions, there are instances in IoT networks where decisions require a more nuanced approach that cannot be easily categorized into discrete options. In these cases, methods like proximal policy optimization or actor-critic techniques could significantly enhance the flexibility and accuracy of IDS solutions. By leveraging the strengths of both policy-based and value-based approaches, IDS can become more adaptable and effective in detecting and mitigating sophisticated cyber threats in IoT networks.

### 7.4. Transitioning from Centralized to Distributed IDS in IoT

Our investigation reveals a significant shift from traditional centralized IDS to more adaptable, distributed models in the IoT domain. This evolution, driven by the integration of edge computing and a focus on resource efficiency, acknowledges the pressing need for IDS solutions that are lightweight yet capable of operating in the stringent confines of IoT systems. This trend highlights the growing demand for IDS frameworks that integrate seamlessly into the diverse and resource-limited IoT landscape, ensuring comprehensive security without sacrificing device efficiency or autonomy. Some studies, such as the implementation of the DQN framework optimized for edge deployment [79] and applying of FL in a distributed IDS architecture [66], exemplify strategic approaches to minimizing IoT devices' computational demands. These approaches address power and computational constraints and enhance data privacy through localized processing, demonstrating the shift towards resource-aware and privacy-preserving distributed learning and IDS strategies. Despite these advancements, distributed IDS approaches remain underrepresented, indicating significant room for innovation in distributed IDS for IoT. The need for IDS solutions that are acutely aware of and designed to overcome the challenges of energy consumption, CPU load, CPU usage, and memory limitations is more critical than ever.

### 7.5. Imitations in Coverage of IoT Application Domains

The papers' analysis demonstrates significant gaps in the coverage of IoT application domains, especially in emerging areas such as Agricultural IoT, Environmental Monitoring, and Retail and Supply Chain IoT. Current research has predominantly focused on more established sectors like Industrial IoT, healthcare, and smart cities, leaving these newer domains underexplored. Addressing these gaps is crucial for providing a comprehensive understanding of DRL-based IDS applications across the full spectrum of IoT.

## 7.6. SDN-IoT and DRL-based IDS Integration

Despite the recognized potential of integrating SDN with IoT through DRL to enhance IDS, researchers show a considerable gap in fully leveraging SDN's capabilities in DRL-based IDS frameworks [66]. This underscores the need for more focused efforts to maximize the benefits of SDN in improving the efficiency of IDS in IoT settings. Challenges in this integration include establishing efficient communication channels between DRL-based IDS agents and SDN controllers for seamless data exchange and decision-making. IoT networks' diverse and expansive nature adds complexity, requiring scalable learning algorithms that remain effective without compromising performance. Additionally, it is crucial to balance the use of SDN-IoT's network management features with the preservation of the decision-making autonomy of DRL-based IDS agents to ensure effective and informed security actions.

## 7.7. Complex Reward Mechanisms

The complexity of IoT security, especially in scenarios ( e.g., safeguarding smart city infrastructures), necessitates advanced DRL-based IDS equipped with adaptive reward mechanisms. These mechanisms must accurately reflect cyber threats' diverse and sophisticated nature. For example, in protecting a smart city from a DDoS attack, a DRL-based IDS benefits from a reward mechanism that considers the criticality of services, potential public safety impacts, and the system's resilience. Such detailed reward structures enable DRL-based IDS to prioritize threats effectively, ensuring critical infrastructure remains protected. This approach enhances the model's decision-making capabilities and ensures continuous adaptation to new threats, maintaining the integrity and safety of smart city IoT ecosystems.

## 7.8. Optimizing Resource Utilization

Integrating DRL into IDS in the IoT ecosystem highlights significant resource management challenges, including CPU load, CPU usage, and energy consumption. This underscores the need for energy-efficient IDS solutions that manage computational demands. Innovative approaches are needed to reduce energy and computational burdens without compromising threat detection accuracy to enhance the scalability and efficiency of DRL-based IDS. Future advancements may include energy-efficient deployment strategies tailored to the power constraints of IoT devices and the implementation of data-efficient learning mechanisms. These efforts are crucial for ensuring robust security measures that comply with the operational limitations of edge computing and the expansive nature of IoT networks.

## 7.9. FL for Privacy

Integrating FL into DRL-based IDS presents a strategic approach to bolstering privacy and security in IoT networks. This method facilitates a collaborative learning environment, enabling devices across the network to contribute to a collective ML model without centralizing sensitive information. Such a decentralized model is beneficial in IoT settings, addressing data privacy issues and managing the vast amounts of data produced. FL ensures data remains localized, reducing risks linked to data transmission and aligning with privacy regulations. For effective deployment of DRL-based IDS, navigating challenges of data diversity, communication efficiency, and synthesizing learning from disparate sources is essential. This technique enhances the capability of edge devices to autonomously detect threats based on local data, contributing to a more intelligent and privacy-conscious IDS framework. In a recent study, [77] emphasized the importance of incorporating privacy-preserving mechanisms, such as differential privacy, within FL frameworks to ensure secure data exchange while minimizing risks of exposure. Their work highlights how these mechanisms, combined with DRL, can enhance the reliability and ethical handling of sensitive IoT data.

## 7.10. Dynamic Threat Detection

The rapidly evolving landscape of IoT security, marked by adversaries devising new attack methods, presents a significant challenge for DRL-based IDS. These systems must recognize known threats and adapt to detect novel, unseen attack vectors. Addressing this challenge requires leveraging transfer learning, enhancing learning algorithms, and employing multi-agent systems with advanced exploration strategies. Transfer learning enables DRL models to apply knowledge from previous tasks to new problems, reducing the training time and data needed for novel threats. Combined with sophisticated learning algorithms, this adaptability enables processing complex data structures and identifying patterns indicative of emerging threats with minimal human intervention. Furthermore, multi-agent systems foster a collaborative learning environment where agents collaborate, enhancing the detection capabilities and robustness of the IDS against evolving threats in the IoT ecosystem.

## 7.11. Real-World Deployment

Transitioning DRL-based IDS from theoretical models and simulations to practical, real-world applications encompasses a range of challenges. Real-world deployment must contend with hardware limitations, network constraints, and the need for real-time processing capabilities. DRL models that perform well in simulations may encounter issues in actual IoT systems, where resource constraints and unpredictable network conditions can impact performance. Addressing these challenges involves optimizing DRL algorithms for efficiency and ensuring they operate in real IoT devices' hardware and network limitations. This includes developing robust models capable of processing and responding to threats in real-time.

## 7.12. Ethical and Legal Considerations

The potential for biases in threat detection, rooted in the training data of ML algorithms, raises serious ethical concerns. Such biases could lead to unfair or incorrect threat assessments, mainly impacting healthcare applications in IoT [113]. Moreover, the automated decision-making capability of DRL-based IDS necessitates a thorough evaluation of its implications. In healthcare, where decisions can directly impact patient well-being and data privacy, the accuracy of these systems is paramount. Errors, such as false positives or negatives, could lead to unnecessary alarms or overlooked threats, each with significant implications [114].

The complexity of DRL-based IDS systems raises concerns about fairness, transparency, and the ethical handling of sensitive data. The work by [71] highlights the necessity of integrating privacy-preserving mechanisms to protect sensitive information during the training and deployment of DRL models. Furthermore, the study emphasizes the critical role of fairness-aware algorithms in mitigating biases inherent in datasets, which can otherwise lead to inequality. By incorporating these approaches, DRL-based IDS enhance their trustworthiness, ensure ethical data handling, and address societal demands for equity and accountability in threat mitigation.

### 7.13. Topology Awareness and Graph Modelization

Graph representation learning and graph RL can significantly enhance IDS capabilities in complex IoT networks, such as those in smart cities. By modeling these networks as graphs, it is possible to leverage the inherent structure to detect anomalies and patterns that indicate potential cyber threats. For instance, identifying unusual communication patterns or critical nodes in the IoT network can help preemptively mitigate risks. Graph-based models enable more efficient allocation of monitoring resources, thereby improving IDS performance in complex environments. While the potential of topology awareness and graph modeling in IDS is evident, it is essential to recognize that this area is still underexplored. Future research should focus on integrating these techniques into DRL-based IDS to develop more sophisticated cybersecurity measures, especially in smart city contexts and other intricate IoT systems.

### 7.14. Evaluating DRL-Based IDS in Live Threat Environments

The lack of studies evaluating DRL-based IDS under real-time cyber threat scenarios presents a significant gap in current research. Testing in live environments is crucial for accurately assessing the system's responsiveness and adaptability to dynamic and unpredictable threats. However, several challenges make such testing difficult. One major challenge is real-world cyber threats' inherent unpredictability and complexity, which vary in form, intensity, and timing. This unpredictability makes it challenging to create controlled yet realistic testing scenarios. Additionally, testing in live environments may expose critical systems to potential harm, raising concerns about the safety and integrity of the IoT ecosystem during the evaluation process. Addressing these challenges is essential for practically implementing DRL-based IDS and ensuring that these systems are robust, reliable, and effective in real-world applications.

### 7.15. Implementing Security Policies

Assume that in a smart city scenario, a DRL-based IDS detects a sophisticated DDoS attack aimed at disrupting traffic management during peak hours. While the IDS identifies the threat, the challenge extends to the post-detection actions to mitigate the attack and preserve the city's infrastructure integrity. Ideally, the IDS should activate a predefined security policy involving rerouting network traffic, isolating affected systems, or temporarily disabling certain functions to contain the attack. In a recent study, [55] proposed a Power-Law-Based Blocking Time Management (PL-BTM) mechanism that dynamically adjusts the blocking time of IP addresses based on their cumulative threat levels. This approach helps defense strategies for specific threat scenarios, ensuring minimal disruption to essential services such as traffic lights and emergency responses while effectively containing malicious activity. However, the lack of research on implementing these automated post-detection responses highlights a significant gap in current IDS capabilities, emphasizing the need for comprehensive security frameworks that include threat detection and effective countermeasures to protect IoT systems against complex cyber threats.

## 8. Conclusion

This systematic review highlights the significant progress in applying DRL-based IDS in IoT systems from 2014 to 2024. DRL methodologies have proven adaptable and effective in enhancing the security of IoT networks, offering improved accuracy and flexibility in detecting and mitigating evolving cyber threats. However, challenges remain, including reliance on outdated datasets, reproducibility issues, and the need for more scalable solutions. Addressing these challenges is essential for further advancing the field. Future research should focus on exploring federated learning, policy learning methods, and the integration of high-level threat intelligence into DRL models, which hold promise for improving the effectiveness and efficiency of IDS solutions.

## References

[1] H. Taherdoost, Security and internet of things: benefits, challenges, and future perspectives, Electronics 12 (8) (2023) 1901.

[2] Z. Li, W. Yao, A two stage lightweight approach for intrusion detection in internet of things, Expert Systems with Applications 257 (2024) 124965.

[3] A. Heidari, M. A. Jabraeil Jamali, Internet of things intrusion detection systems: a comprehensive review and future directions, Cluster Computing 26 (6) (2023) 3753–3780.

[4] S. Santhosh Kumar, M. Selvi, A. Kannan, A comprehensive survey on machine learning-based intrusion detection systems for secure communication in internet of things, Computational Intelligence and Neuroscience 2023 (1) (2023) 8981988.

[5] D. Vamvakas, P. Michailidis, C. Korkas, E. Kosmatopoulos, Review and evaluation of reinforcement learning frameworks on smart grid applications, Energies 16 (14) (2023) 5326.

[6] M. A. Merzouk, C. Neal, J. Delas, R. Yaich, N. Boulahia-Cuppens, F. Cuppens, Adversarial robustness of deep reinforcement learning-based intrusion detection, International Journal of Information Security (2024) 1–27.

[7] A. Gueriani, H. Kheddar, A. C. Mazari, Deep reinforcement learning for intrusion detection in iot: A survey, in: 2023 2nd International Conference on Electronics, Energy and Measurement (IC2EM), Vol. 1, IEEE, 2023, pp. 1–7.

[8] S. E. Li, Deep reinforcement learning, in: Reinforcement Learning for Sequential Decision and Optimal Control, Springer, 2023, pp. 365–402.

[9] X. Wang, S. Wang, X. Liang, D. Zhao, J. Huang, X. Xu, B. Dai, Q. Miao, Deep reinforcement learning: a survey, IEEE Transactions on Neural Networks and Learning Systems (2022).

[10] I. H. Sarker, A. I. Khan, Y. B. Abushark, F. Alsolami, Internet of things (iot) security intelligence: a comprehensive overview, machine learning solutions and research directions, Mobile Networks and Applications 28 (1) (2023) 296–312.

[11] O. H. Abdulganiyu, T. Ait Tchakoucht, Y. K. Saheed, A systematic literature review for network intrusion detection system (ids), International journal of information security 22 (5) (2023) 1125–1162.

[12] A. M. K. Adawadkar, N. Kulkarni, Cyber-security and reinforcement learning—a brief survey, Engineering Applications of Artificial Intelligence 114 (2022) 105116.

[13] J. F. Cevallos M, A. Rizzardi, S. Sicari, A. Coen Porisini, Deep reinforcement learning for intrusion detection in internet of things:: Best practices, lessons learnt, and open challenges (2023).

[14] W. Chen, X. Qiu, T. Cai, H.-N. Dai, Z. Zheng, Y. Zhang, Deep reinforcement learning for internet of things: A comprehensive survey, IEEE Communications Surveys & Tutorials 23 (3) (2021) 1659–1692.

[15] A. Rizzardi, S. Sicari, A. C. Porisini, et al., Deep reinforcement learning for intrusion detection in internet of things: Best practices, lessons learnt, and open challenges, Computer Networks 236 (2023) 110016.

[16] K. Arshad, R. F. Ali, A. Muneer, I. A. Aziz, S. Naseer, N. S. Khan, S. M. Taib, Deep reinforcement learning for anomaly detection: A systematic review, IEEE Access 10 (2022) 124017–124035.

[17] M. Humayun, N. Tariq, M. Alfayad, M. Zakwan, G. Alwakid, M. Assiri, Securing the internet of things in artificial intelligence era: A comprehensive survey, IEEE Access (2024).

[18] M. Ozkan-Ozay, E. Akin, Ö. Aslan, S. Kosunalp, T. Iliev, I. Stoyanov, I. Beloev, A comprehensive survey: Evaluating the efficiency of artificial intelligence and machine learning techniques on cyber security solutions, IEEE Access (2024).

[19] A. Dhar Dwivedi, R. Singh, K. Kaushik, R. Rao Mukkamala, W. S. Alnumay, Blockchain and artificial intelligence for 5g-enabled internet of things: Challenges, opportunities, and solutions, Transactions on Emerging Telecommunications Technologies 35 (4) (2024) e4329.

[20] S. Keele, et al., Guidelines for performing systematic literature reviews in software engineering (2007).

[21] W. AlAqqad, M. Nijim, U. Onyeakazi, H. Albataineh, Cyber edge: Mitigating cyber-attacks in edge computing using intrusion detection system, in: International Conference on Advances in Computing Research, Springer, 2024, pp. 292–305.

[22] A. Nag, M. M. Hassan, A. Das, A. Sinha, N. Chand, A. Kar, V. Sharma, A. Alkhayyat, Exploring the applications and security threats of internet of thing in the cloud computing paradigm: A comprehensive study on the cloud of things, Transactions on Emerging Telecommunications Technologies 35 (4) (2024) e4897.

[23] F. A. Alaba, Iot architecture layers, in: Internet of Things: A Case Study in Africa, Springer, 2024, pp. 65–85.

[24] M. Adam, M. Hammoudeh, R. Alrawashdeh, B. Alsulaimy, A survey on security, privacy, trust, and architectural challenges in iot systems, IEEE Access (2024).

[25] O. Arshi, A. Rai, G. Gupta, J. K. Pandey, S. Mondal, Iot in energy: a comprehensive review of technologies, applications, and future directions, Peer-to-Peer Networking and Applications (2024) 1–40.

[26] E. N. Amachaghi, M. Shojafar, C. H. Foh, K. Moessner, A survey for intrusion detection systems in open ran, IEEE Access (2024).

[27] S. J. Soheli, N. Jahan, M. B. Hossain, A. Adhikary, A. R. Khan, M. Wahiduzzaman, Smart greenhouse monitoring system using internet of things and artificial intelligence, Wireless Personal Communications 124 (4) (2022) 3603–3634.

[28] M. Ahmid, O. Kazar, A comprehensive review of the internet of things security, Journal of Applied Security Research 18 (3) (2023) 289–305.

[29] S. N. G. Aryavalli, H. Kumar, Top 12 layer-wise security challenges and a secure architectural solution for internet of things, Computers and Electrical Engineering 105 (2023) 108487.

[30] V. V. Vegesna, Methodology for mitigating the security issues and challenges in the internet of things (iot) framework for enhanced security, Asian Journal of Basic Science & Research 5 (1) (2023) 85–102.

[31] B. Kaur, S. Dadkhah, F. Shoeleh, E. C. P. Neto, P. Xiong, S. Iqbal, P. Lamontagne, S. Ray, A. A. Ghorbani, Internet of things (iot) security dataset evolution: Challenges and future directions, Internet of Things 22 (2023) 100780.

[32] M. A. Jamshed, K. Ali, Q. H. Abbasi, M. A. Imran, M. Ur-Rehman, Challenges, applications, and future of wireless sensors in internet of things: A review, IEEE Sensors Journal 22 (6) (2022) 5482–5494.

[33] M. Agoramoorthy, A. Ali, D. Sujatha, M. R. TF, G. Ramesh, An analysis of signature-based components in hybrid intrusion detection systems, in: 2023 Intelligent Computing and Control for Engineering and Business Systems (ICCEBS), IEEE, 2023, pp. 1–5.

[34] D. Ramakrishna, M. A. Shaik, A comprehensive analysis of cryptographic algorithms: Evaluating security, efficiency, and future challenges, IEEE Access (2024).

[35] J. M. Kizza, System intrusion detection and prevention, in: Guide to computer network security, Springer, 2024, pp. 295–323.

[36] S. Chell, S. Chakare, P. Sohan, S. Sandosh, Real-time threat detection and mitigation in web api development, in: 2024 International Conference on Electrical Electronics and Computing Technologies (ICEECT), Vol. 1, IEEE, 2024, pp. 1–9.

[37] P. Bajaj, S. Mishra, A. Paul, Comparative analysis of stack-ensemble-based intrusion detection system for single-layer and cross-layer dos attack detection in iot, SN Computer Science 4 (5) (2023) 562.

[38] K. Lian, L. Zhang, G. Yang, S. Mao, X. Wang, Y. Zhang, M. Yang, Component security ten years later: An empirical study of cross-layer threats in real-world mobile applications, Proceedings of the ACM on Software Engineering 1 (FSE) (2024) 70–91.

[39] H. Azzaoui, A. Z. E. Boukhamla, P. Perazzo, M. Alazab, V. Ravi, A lightweight cooperative intrusion detection system for rpl-based iot, Wireless Personal Communications 134 (4) (2024) 2235–2258.

[40] R. Ahmad, I. Alsmadi, W. Alhamdani, L. Tawalbeh, A comprehensive deep learning benchmark for iot ids, Computers & Security 114 (2022) 102588.

[41] S. A. Abdulkareem, C. H. Foh, M. Shojafar, F. Carrez, K. Moessner, Network intrusion detection: An iot and non iot-related survey, IEEE Access (2024).

[42] A. A. Anitha, L. Arockiam, A review on intrusion detection systems to secure iot networks, International Journal of Computer Networks and Applications 9 (1) (2022) 38–50.

[43] K. Vaigandla, N. Azmi, R. Karne, Investigation on intrusion detection systems (idss) in iot, International Journal of Emerging Trends in Engineering Research 10 (3) (2022).

[44] X. Feng, J. Han, R. Zhang, S. Xu, H. Xia, Security defense strategy algorithm for internet of things based on deep reinforcement learning, High-Confidence Computing 4 (1) (2024) 100167.

[45] A. K. Shukla, Detection of anomaly intrusion utilizing self-adaptive grasshopper optimization algorithm, Neural Computing and Applications 33 (13) (2021) 7541–7561.

[46] D. K. Dake, J. D. Gadze, G. S. Klogo, H. Nunoo-Mensah, Multi-agent reinforcement learning framework in sdn-iot for transient load detection and prevention, Technologies 9 (3) (2021) 44.

[47] S. Janakiraman, M. Deva Priya, A deep reinforcement learning-based ddos attack mitigation scheme for securing big data in fog-assisted cloud environment, Wireless Personal Communications 130 (4) (2023) 2869–2886.

[48] A. Pashaei, M. E. Akbari, M. Z. Lighvan, A. Charmin, Early intrusion detection system using honeypot for industrial control networks, Results in Engineering 16 (2022) 100576.

[49] Elsevier, Engineering village: Compendex, https://www.engineeringvillage.com/search/doc/compendex, accessed on: June 5, 2024.

[50] C. Analytics, Web of science, https://www.webofscience.com, accessed on: June 5, 2024.

[51] M. J. Page, J. E. McKenzie, P. M. Bossuyt, I. Boutron, T. C. Hoffmann, C. D. Mulrow, L. Shamseer, J. M. Tetzlaff, E. A. Akl, S. E. Brennan, et al., The prisma 2020 statement: an updated guideline for reporting systematic reviews, International journal of surgery 88 (2021) 105906.

[52] A. Carrera-Rivera, W. Ochoa, F. Larrinaga, G. Lasa, How-to conduct a systematic literature review: A quick guide for computer science research, MethodsX 9 (2022) 101895.

[53] Rayyan – intelligent systematic review, https://www.rayyan.ai/.

[54] A. Rejeb, K. Rejeb, H. Treiblmaier, A. Appolloni, S. Alghamdi, Y. Alhasawi, M. Iranmanesh, The internet of things (iot) in healthcare: Taking stock and moving forward, Internet of Things 22 (2023) 100721.

[55] Y. Liu, K.-F. Tsang, C. K. Wu, Y. Wei, H. Wang, H. Zhu, Ieee p2668-compliant multi-layer iot-ddos defense system using deep reinforcement learning, IEEE Transactions on Consumer Electronics 69 (1) (2022) 49–64.

[56] X. Liu, W. Yu, F. Liang, D. Griffith, N. Golmie, On deep reinforcement learning security for industrial internet of things, Computer Communications 168 (2021) 20–32.

[57] T. Ramana, M. Thirunavukkarasan, A. S. Mohammed, G. G. Devarajan, S. M. Nagarajan, Ambient intelligence approach: Internet of things based decision performance analysis for intrusion detection, Computer Communications 195 (2022) 315–322.

[58] S. Vadigi, K. Sethi, D. Mohanty, S. P. Das, P. Bera, Federated reinforcement learning based intrusion detection system using dynamic attention mechanism, Journal of Information Security and Applications 78 (2023) 103608.

[59] X. Ma, Y. Li, Y. Gao, Decision model of intrusion response based on markov game in fog computing environment, Wireless Networks 29 (8) (2023) 3383–3392.

[60] R. Zhang, H. Xia, C. Liu, R.-b. Jiang, X.-g. Cheng, Anti-attack scheme for edge devices based on deep reinforcement learning, Wireless Communications and Mobile Computing 2021 (2021) 1–9.

[61] M. Al-Fawa'reh, J. Abu-Khalaf, P. Szewczyk, J. J. Kang, Malbot-drl: Malware botnet detection using deep reinforcement learning in iot networks, IEEE Internet of Things Journal (2023).

[62] M. Lopez-Martin, B. Carro, A. Sanchez-Esguevillas, Application of deep reinforcement learning to intrusion detection for supervised problems, Expert Systems with Applications 141 (2020) 112963.

[63] M. M. Almasri, A. M. Alajlan, A novel-cascaded anfis-based deep reinforcement learning for the detection of attack in cloud iot-based smart city applications, Concurrency and Computation: Practice and Experience 35 (22) (2023) e7738.

[64] J. Wang, Y. Liu, W. Zhang, X. Yan, N. Zhou, Z. Jiang, Relfa: Resist link flooding attacks via renyi entropy and deep reinforcement learning in sdn-iot, China Communications 19 (7) (2022) 157–171.

[65] N. Kandhoul, S. K. Dhurandher, Deep q learning based secure routing approach for oppiot networks, Internet of Things 20 (2022) 100597.

[66] J. Wang, J. Liu, Deep learning for securing software-defined industrial internet of things: attacks and countermeasures, IEEE Internet of Things Journal 9 (13) (2021) 11179–11189.

[67] K. Ren, Y. Zeng, Y. Zhong, B. Sheng, Y. Zhang, Mafsids: a reinforcement learning-based intrusion detection model for multi-agent feature selection networks, Journal of Big Data 10 (1) (2023) 137.

[68] B. Yang, M. H. Arshad, Q. Zhao, Packet-level and flow-level network intrusion detection based on reinforcement learning and adversarial training, Algorithms 15 (12) (2022) 453.

[69] Y. Feng, W. Zhang, S. Yin, H. Tang, Y. Xiang, Y. Zhang, A collaborative stealthy ddos detection method based on reinforcement learning at the edge of the internet of things, IEEE Internet of Things Journal (2023).

[70] H. Karthikeyan, G. Usha, Real-time ddos flooding attack detection in intelligent transportation systems, Computers and Electrical Engineering 101 (2022) 107995.

[71] F. Mesadieu, D. Torre, A. Chennameneni, Leveraging deep reinforcement learning technique for intrusion detection in scada infrastructure, IEEE Access (2024).

[72] H. H. R. Sherazi, R. Iqbal, F. Ahmad, Z. A. Khan, M. H. Chaudary, Ddos attack detection: A key enabler for sustainable communication in internet of vehicles, Sustainable Computing: Informatics and Systems 23 (2019) 13–20.

[73] R. Heartfield, G. Loukas, A. Bezemskij, E. Panaousis, Self-configurable cyber-physical intrusion detection for smart homes using reinforcement learning, IEEE Transactions on Information Forensics and Security 16 (2020) 1720–1735.

[74] P. Radoglou-Grammatikis, K. Rompolos, P. Sarigiannidis, V. Argyriou, T. Lagkas, A. Sarigiannidis, S. Goudos, S. Wan, Modeling, detecting, and mitigating threats against industrial healthcare systems: a combined software defined networking and reinforcement learning approach, IEEE Transactions on Industrial Informatics 18 (3) (2021) 2041–2052.

[75] C. Hu, J. Yan, X. Liu, Reinforcement learning-based adaptive feature boosting for smart grid intrusion detection, IEEE Transactions on Smart Grid (2022).

[76] L. Nie, W. Sun, S. Wang, Z. Ning, J. J. Rodrigues, Y. Wu, S. Li, Intrusion detection in green internet of things: a deep deterministic policy gradient-based algorithm, IEEE Transactions on Green Communications and Networking 5 (2) (2021) 778–788.

[77] N. M. Al-Maslamani, B. S. Ciftler, M. Abdallah, M. M. Mahmoud, Toward secure federated learning for iot using drl-enabled reputation mechanism, IEEE Internet of Things Journal 9 (21) (2022) 21971–21983.

[78] J. Parras, A. Almodóvar, P. A. Apellániz, S. Zazo, Inverse reinforcement learning: a new framework to mitigate an intelligent backoff attack, IEEE Internet of Things Journal 9 (24) (2022) 24790–24799.

[79] S. Tharewal, M. W. Ashfaque, S. S. Banu, P. Uma, S. M. Hassen, M. Shabaz, Intrusion detection system for industrial internet of things based on deep reinforcement learning, Wireless Communications and Mobile Computing 2022 (2022) 1–8.

[80] N. S. Alotaibi, H. I. Ahmed, S. O. M. Kamel, Dynamic adaptation attack detection model for a distributed multi-access edge computing smart city, Sensors 23 (16) (2023) 7135.

[81] S. Dong, Y. Xia, T. Peng, Network abnormal traffic detection model based on semi-supervised deep reinforcement learning, IEEE Transactions on Network and Service Management 18 (4) (2021) 4197–4212.

[82] M. Alauthman, N. Aslam, M. Al-Kasassbeh, S. Khan, A. Al-Qerem, K.-K. R. Choo, An efficient reinforcement learning-based botnet detection approach, Journal of Network and Computer Applications 150 (2020) 102479.

[83] V. Praveena, A. Vijayaraj, P. Chinnasamy, I. Ali, R. Alroobaea, S. Y. Alyahyan, M. A. Raza, Optimal deep reinforcement learning for intrusion detection in uavs, Computers, Materials & Continua 70 (2) (2022) 2639–2653.

[84] V. Juneja, S. K. Dinkar, D. V. Gupta, An anomalous co-operative trust & pg-drl based vampire attack detection & routing, Concurrency and Computation: Practice and Experience 34 (3) (2022) e6557.

[85] H. Nandanwar, R. Katarya, Deep learning enabled intrusion detection system for industrial iot environment, Expert Systems with Applications 249 (2024) 123808.

[86] Intrusion detection evaluation dataset (iscxids2012), https://www.unb.ca/cic/datasets/ids.html.

[87] Unsw-nb15 dataset, https://research.unsw.edu.au/projects/unsw-nb15-dataset.

[88] Intrusion detection evaluation dataset (cic-ids2017), https://www.unb.ca/cic/datasets/ids-2017.html.

[89] Industrial internet of things (iiot) research 2018, https://www.cse.wustl.edu/~jain/iiot/index.html.

[90] Advanced industrial internet of things (iiot) research 2021, https://www.cse.wustl.edu/~jain/iiot2/index.html.

[91] N-baiot dataset, https://www.kaggle.com/datasets/mkashifn/nbaiot-dataset/code, accessed: 2024-05-07.

[92] Hogzilla ids dataset, https://ids-hogzilla.org/dataset/.

[93] Iotid20 dataset, https://www.kaggle.com/datasets/rohulaminlabid/iotid20-dataset.

[94] P. Radoglou-Grammatikis, K. Rompolos, T. Lagkas, V. Argyriou, P. Sarigiannidis, Iec 60870-5-104 intrusion detection dataset (2022). doi:10.21227/fj7s-f281.
URL https://dx.doi.org/10.21227/fj7s-f281

[95] Awid intrusion detection dataset, https://github.com/Bee-Mar/AWID-Intrusion-Detection.

[96] Bot-iot dataset, https://research.unsw.edu.au/projects/bot-iot-dataset.

[97] Ddos 2019 dataset, https://www.unb.ca/cic/datasets/ddos-2019.html.

[98] Medbiot dataset, https://cs.taltech.ee/research/data/medbiot/.

[99] U. Singh, M. Rizwan, Scada system dataset exploration and machine learning based forecast for wind turbines, Results in Engineering 16 (2022) 100640.

[100] J. F. Cevallos Moreno, A. Rizzardi, S. Sicari, A. Coen-Porisini, Deep reinforcement learning for intrusion detection in internet of things: Best practices, lessons learnt, and open challenges, Lessons Learnt, and Open Challenges.

[101] A. K. Zamani, A. Chapnevis, Botnet intrusion detection system in internet of things with developed deep learning, arXiv preprint arXiv:2207.04503 (2022).

[102] H. Kheddar, D. W. Dawoud, A. I. Awad, Y. Himeur, M. K. Khan, Reinforcement-learning-based intrusion detection in communication networks: A review, IEEE Communications Surveys & Tutorials (2024).

[103] S. Anbazhagan, R. Mugelan, Next-gen resource optimization in nb-iot networks: Harnessing soft actor–critic reinforcement learning, Computer Networks 252 (2024) 110670.

[104] S. Yang, L. Zhuang, J. Zhang, J. Lan, B. Li, A multi-policy deep reinforcement learning approach for multi-objective joint routing and scheduling in deterministic networks, IEEE Internet of Things Journal (2024).

[105] Kdd cup 1999 data, https://www.kaggle.com/datasets/galaxyh/kdd-cup-1999-data (1999).

[106] Edge iiotset pre-processing, https://www.kaggle.com/code/mohamedamineferrag/edge-iiotset-pre-processing.

[107] U. of New Brunswick, Iot dataset 2023, accessed: 2025-01-09 (2023).
URL https://www.unb.ca/cic/datasets/iotdataset-2023.html

[108] B. Saeed, S. Arshad, S. M. U. Saeed, M. A. Azam, Enhancing iot security: Federated learning with gans for effective attack detection, in: 2023 20th International Bhurban Conference on Applied Sciences and Technology (IBCAST), IEEE, 2023, pp. 570–575.

[109] ns 3 Project, ns-3 network simulator.
URL https://www.nsnam.org/

[110] C. O. Project, An introduction to cooja simulator.
URL https://github.com/contiki-os/contiki/wiki/An-Introduction-to-COOJA

[111] O. Project, Omnet++ network simulator.
URL https://omnetpp.org/

[112] M. Adam, U. Baroud, Federated learning for iot: Applications, trends, taxonomy, challenges, current solutions, and future directions, IEEE Open Journal of the Communications Society (2024).

[113] F. Chen, L. Wang, J. Hong, J. Jiang, L. Zhou, Unmasking bias in artificial intelligence: a systematic review of bias detection and mitigation strategies in electronic health record-based models, Journal of the American Medical Informatics Association 31 (5) (2024) 1172–1183.

[114] M. Tabassum, S. Mahmood, A. Bukhari, B. Alshemaimri, A. Daud, F. Khalique, Anomaly-based threat detection in smart health using machine learning, BMC Medical Informatics and Decision Making 24 (1) (2024) 347.

# Appendix

Table 6: Abbreviations used in this research.

| Abbreviation | Meaning |
|---|---|
| AC | Actor-Critic |
| AFB | Adaptive Feature Boosting |
| AI | Artificial Intelligence |
| APT | Advanced Persistent Threat |
| CRN | Cognitive Radio Network |
| DDPG | Deep Deterministic Policy Gradient |
| DDQN | Double Deep Q-Network |
| DL | Deep Learning |
| DoS | Denial of Service |
| DQL | Deep Q-Learning |
| DRL | Deep Reinforcement Learning |
| DQN | Deep Q-Network |
| FL | Federated Learning |
| GAN | Generative Adversarial Network |
| GCN | Graph Convolutional Network |
| GOA | Grasshopper Optimization Algorithm |
| IDS | Intrusion Detection System |
| IEC | International Electrotechnical Commission |
| IoT | Internet of Things |
| IRL | Inverse Reinforcement Learning |
| ITS | Intelligent Transportation Systems |
| LFA | Link Flooding Attack |
| MADDPG | Multi-Agent Deep Deterministic Policy Gradient |
| MAB | Multi-Armed Bandit |
| ML | Machine Learning |
| MITM | Man-in-the-Middle |
| NFQ | Neural Fitted Q-Iteration |
| NSL-KDD | Network Security Lab Knowledge Discovery Database |
| PG-DRL | Policy Gradient Deep Reinforcement Learning |

| Abbreviation | Meaning |
|---|---|
| PPO | Proximal Policy Optimization |
| RL | Reinforcement Learning |
| SAC | Soft Actor-Critic |
| SARSA | State-Action-Reward-State-Action |
| SCADA | Supervisory Control and Data Acquisition |
| SDN | Software-Defined Networking |
| UAV | Unmanned Aerial Vehicles |
| V2I | Vehicle-to-Infrastructure |
| V2V | Vehicle-to-Vehicle |
| V2X | Vehicle-to-Everything |
| WSN | Wireless Sensor Network |
| R2L | Remote-to-Local |
| U2R | User-to-Root |
| AUC | Area Under the Curve |
| FPR | False Positive Rate |
| TPR | True Positive Rate |
| IoV | Internet of Vehicles |
| IIoT | Industrial Internet of Things |
| QoS | Quality of Service |
| HIL | Hardware-in-the-Loop |
| kNN | k-Nearest Neighbors |
| DTW | Dynamic Time Warping |
| SVM | Support Vector Machine |
| DT | Decision Tree |
| RFC | Random Forest Classifier |
| CAML | Convolutional Attention Memory Network |
| RSU | Roadside Units |
| IDPS | Intrusion Detection and Prevention System |
| TRE | Time Relative Error |
| DSR | Dynamic Source Routing |